\documentclass[a4paper]{article}
\usepackage{a4wide}
\usepackage{graphicx,xcolor}
\graphicspath{{./images/}}
\usepackage{caption}
\usepackage{subcaption}
\usepackage{empheq}
\usepackage{amsthm}
\usepackage{amssymb, latexsym, mathrsfs}
\usepackage{amsmath}
\usepackage{dsfont}
\usepackage{enumerate}
\usepackage{algorithm}
\usepackage{color}
\usepackage{transparent}
\usepackage{url}
\usepackage[affil-it]{authblk}
\usepackage{booktabs}

\usepackage[utf8]{inputenc}
\usepackage[T1]{fontenc}
\usepackage{lmodern}

\newtheorem{rmk}{Remark}

\usepackage[american,cuteinductors,smartlabels]{circuitikz}
\usepackage{tikz}
\usepackage{schemabloc}
\usetikzlibrary{shapes,arrows}

\begin{document}

\title{Plug-and-play and coordinated control for bus-connected AC islanded microgrids}

\author[1]{Stefano Riverso 
       \thanks{Electronic address: \texttt{riverss@utrc.utc.com}} }
\author[2]{Michele Tucci 
\thanks{Electronic address:
  			\texttt{michele.tucci02@universitadipavia.it}}}
\author[3]{Juan C. Vasquez
       \thanks{Electronic address: \texttt{juq@et.aau.dk}} }
\author[3]{Josep M. Guerrero
       \thanks{Electronic address: \texttt{joz@et.aau.dk}} }
\author[4]{Giancarlo Ferrari-Trecate
       		\thanks{Electronic address: \texttt{giancarlo.ferraritrecate@epfl.ch}}}

\affil[1]{United Technologies Research Center Ireland, 4th Floor, Penrose Business Center, Penrose Wharf, Cork, Ireland}
\affil[2]{Dipartimento di Ingegneria Industriale e dell'Informazione, Universit\`a degli Studi di Pavia, Italy}
\affil[3]{Institute of Energy Technology, Aalborg University,
	9220 Aalborg, Denmark }
\affil[4]{Automatic Control Laboratory,
	\'Ecole Polytechnique F\'ed\'erale de Lausanne (EPFL), Switzerland
	}
\date{\textbf{Technical Report}\\ March, 2017}

\maketitle

\begin{abstract}
This paper presents a distributed control architecture for voltage and frequency stabilization in AC islanded microgrids. In the primary control layer, each generation unit is equipped with a local controller acting on the corresponding voltage-source converter. Following the plug-and-play design approach previously proposed by some of the authors, whenever the addition/removal of a distributed generation unit is required, feasibility of the operation is automatically checked by designing local controllers through convex optimization. The update of the voltage-control layer, when units plug -in/-out, is therefore automatized and stability of the microgrid is always preserved. Moreover, local control design is based only on the knowledge of parameters of power lines and it does not require to store a global microgrid model. In this work, we focus on bus-connected microgrid topologies and enhance the primary plug-and-play layer with local virtual impedance loops and secondary coordinated controllers ensuring bus voltage tracking and reactive power sharing. In particular, the secondary control architecture is distributed, hence mirroring the modularity of the primary control layer. We validate primary and secondary controllers by performing experiments with balanced, unbalanced and nonlinear loads, on a setup composed of three bus-connected distributed generation units. Most importantly, the stability of the microgrid after the addition/removal of distributed generation units is assessed. Overall, the experimental results show the feasibility of the proposed modular control design framework, where generation units can be added/removed on the fly, thus enabling the deployment of virtual power plants that can be resized over time.

\textit{Keywords:} Distributed control, secondary control, plug-and-play design, AC islanded
       microgrid, voltage and frequency control.
\end{abstract}

\newpage

 \section{Introduction}
Islanded microGrids (ImGs) are autonomous energy systems composed of the interconnection of Distributed Generation Units (DGUs) and loads. In view of their capability of supplying loads in absence of a connection to the main grid, ImGs provide a flexible solution for bringing power to remote areas or islands \cite{riverso2015plug, Olivares2014, Guerrero2013,ranaboldo2014renewable,kyriakarakos2011polygeneration}. Furthermore, they can be used for improving the robustness of the main grid, e.g. by guaranteeing power supply after the occurrence of an islanding event. Self-sufficient and flexible generation islands also promote the deregulation of the energy market and they have been advocated as a key component of future smart power systems \cite{kumagai2014rise}. One of the key issues of ImGs is scalability, i.e. how to add and remove DGUs without compromising the safe operation of the whole system. This problem is not trivial, as voltage stability must be guaranteed by regulating the Voltage-Source Converter (VSC) of individual DGUs. Furthermore, in order to make online computations grow nicely with the ImG size, decentralized control architectures (where each DGU is equipped with a local regulator) must be used \cite{Etemadi2012a,Babazadeh2013,7798653}. As each local controller measures variables of the corresponding DGU only, the plugging-in or -out of a DGU can easily destabilize the whole ImG if the control layer is not properly updated \cite{Riverso2014c}.
In order to overcome this problem, in \cite{riverso2015plug,Riverso2014c} the authors present a methodology for designing local primary regulators in order to allow Plug-and-Play (PnP) operations \cite{Riverso2013c,Riverso2014} while preserving voltage and frequency stability of the overall ImG. More precisely, when a DGU issues a plug-in or -out request, the safety of the operation is first checked by solving optimization problems that involve only the corresponding DGU and parameters of power lines connecting it to the neighboring units. If the test is passed, stability-preserving regulators are synthesized for the DGU and its neighbors. Compared to other decentralized primary control architectures for ImGs, such as the combination of inner loops and droop controllers \cite{Guerrero2013,de2007voltage,kim2011mode},
PnP controllers are simpler and they are not based on the idea of mimicking regulators for inertial generators. Moreover, differently from decentralized regulators based on robust control \cite{Etemadi2012a,Babazadeh2013}, the knowledge of the whole ImG model is not required for local control design. Rather, just the parameters of power lines are used in the synthesis algorithm.

In this paper, we adapt PnP controllers proposed in \cite{riverso2015plug, Riverso2014c} for achieving several goals and we validate the control architecture through experiments performed on an AC ImG facility. First, since the method in \cite{riverso2015plug} allows to compute local stabilizing controllers for ImGs arranged in \textit{load-connected} topologies only (i.e. where local loads appear at the output terminals of each DGU), we propose a simple procedure for mapping \textit{bus-connected} ImGs (i.e. networks with a common load, supplied by all the DGUs - which are frequently found in several applications) into their equivalent load-connected models. Secondly, we enhance basic PnP controllers with virtual impedance loops, which are needed for setting the output impedance of bus-connected VSCs. Thirdly, we equip the PnP-controlled ImG with a secondary coordinated control layer for voltage tracking at the load bus and sharing of reactive power among parallel VSCs. Notably, since secondary controllers are commonly used in ImGs \cite{Etemadi2012a}, we aim to show how they can be easily coupled with low-level PnP controllers for improving performance without compromising the collective ImG stability.

We conduct several experiments on a bus-connected ImG and assess performance of PnP controllers in tracking set-point voltages. Besides showing the negligible impact of transients when the addition/removal of a DGU is performed, we also demonstrate the good behavior of the closed-loop ImG in presence of both linear and nonlinear loads. 

The paper is organized as follows. In Section \ref{sec:ImG_models}, we recall the electrical model of a load-connected ImG and derive the formulae for computing the equivalent load-connected network of a bus-connected one. In Section \ref{sec:PnP_primary}, we briefly  describe the PnP control approach in \cite{riverso2015plug} and show how to extend the design of PnP voltage and frequency regulators to the case of bus-connected ImGs. In Section \ref{sec:coordinatedControl}, the secondary coordinated control layer is presented. Section \ref{sec:exp_res} is devoted to the assessment, through experiments, of PnP control alone and in combination with the secondary coordinated control layer. Some conclusions are presented in Section \ref{sec:conclusions}.

  \section{Microgrid models}
\label{sec:ImG_models}
\subsection{Load-connected DGU model}
\begin{figure}
	\centering
	\includegraphics[scale=1]{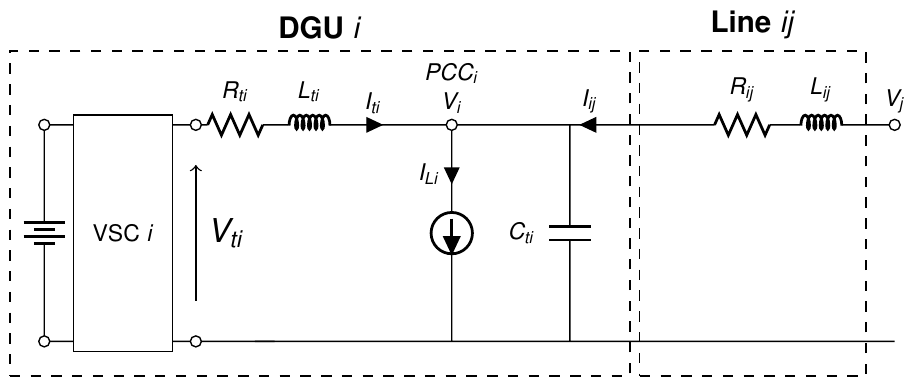}
	\caption{Equivalent single-phase electrical scheme of DGU $i$ and power line $ij$ in a load-connected ImG.}
	\label{fig:load_connected_model}
\end{figure}

In this Section, we present the electrical ImG model considered in \cite{riverso2015plug} for designing decentralized voltage and frequency PnP regulators. We assume three-phase electrical signals without zero-sequence
components and balanced network parameters. The single-phase equivalent scheme of DGU $i$ is shown in Figure \ref{fig:load_connected_model}. As in \cite{riverso2015plug}, we have a DC
voltage source for modeling a generic renewable resource, and
a VSC is controlled in order to supply a
\textit{local load}  (hence the name load-connected topology) connected to the Point of Common Coupling (PCC) through an
$RLC$ filter. Moreover, we assume that
loads $I_{Li}$ are unknown and we treat them as exogenous disturbances \cite{Babazadeh2013}.

Assume now that our ImG is composed of $N$ DGUs. We can
define the set $\mathcal{V}_{DGU}=\{1,\dots,N\}$, and call two DGUs \textit{neighbors} if there is a
power line connecting their PCCs. In particular, we denote the subset of neighbors of DGU
$i$ with $\mathcal{N}_i\subset\mathcal{V}_{DGU}$. We further observe that the neighboring relation is
symmetric (i.e. $j\in\mathcal{N}_i$ implies
$i\in\mathcal{N}_j$).

Let $\omega_0$ be the reference network frequency. By exploiting Quasi Stationary Line (QSL) approximation of power lines \cite{riverso2015plug,Venkatasubramanian1995}, we get the following DGU model in $dq$ reference frame
(rotating with speed $\omega_0$) \cite{riverso2015plug,TucciFloriduz_Kron}
\begin{subequations}
\label{DGUeqx}
 \begin{empheq}[left=$DGU \emph{i}:\quad$\empheqlbrace]{align}
\label{eq:LC_model_1}\frac{\mathrm d}{\mathrm dt} V_i^{dq}&=-\mathrm i \omega_0
V_i^{dq}+
\frac{I_{ti}^{dq}}{C_{ti}}-\frac{I_{Li}^{dq}}{C_{ti}} + \sum_{j\in\mathcal{N}_i} \frac{1}{C_{ti}}\left(\frac{V_j^{dq}-V_i^{dq}}{Z_{ij}}\right)\\
\label{eq:LC_model_2}\frac{\mathrm d}{\mathrm dt}I_{ti}^{dq} &=-\left( \frac{R_{ti}}{L_{ti}}+\mathrm i \omega_0 \right) I_{ti}^{dq} - \frac{V_i^{dq}}{L_{ti}} + \frac{V_{ti}^{dq}}{L_{ti}} 
\end{empheq}
\end{subequations}
where quantities $V_{i}$, $V_{j}$, $I_{ti}$, $I_{Li}$, $V_{ti}$, $R_{ti}$, $C_{ti}$ and $L_{ti}$ are shown in Figure \ref{fig:load_connected_model}. Notably, (i) $V_i$ and $I_{ti}$ represent the $i$-th PCC voltage and
filter current, respectively, (ii) $V_{ti}$ is the command input to the corresponding
VSC, (iii) $R_{ti}$, $L_{ti}$ and $C_{ti}$ are the converter parameters, and (iv) $V_{j}$ indicates the voltage at the
PCC of each neighboring DGU $j\in\mathcal{N}_i$. Moreover, $Z_{ij} = R_{ij} + \mathrm{i}\omega_0 L_{ij}$, where $R_{ij}$ and $L_{ij}$ are, respectively, the resistance
and impedance of the three-phase power line connecting DGUs $i$ and $j$. 
Finally, we indicate with $\mathcal{G}_{lc} = \left(\mathcal{V}_{DGU}, \mathcal{E}_{lc}, W_{lc} \right)$ the directed electric graph associated to the load-connected ImG (see, e.g., Figure \ref{fig:Y_Delta}-b), where weights $W_{ij}$ of edges $e_{ij}\in\mathcal{E}_{lc}$ coincide with the admittances $\frac{1}{Z_{ij}}$ in \eqref{eq:LC_model_1}.

\subsection{From bus-connected to load-connected topology}
\label{sec:load-connected_to_bus-connected}
\begin{figure}
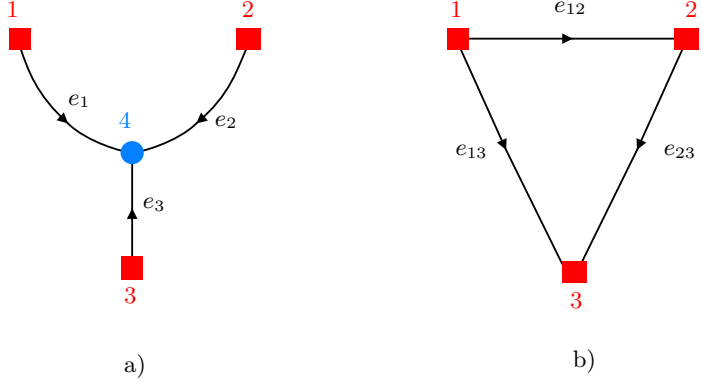

	\small
	\centering
	\begin{pgfpicture}{0cm}{0cm}{288pt}{148pt}
\pgfsetxvec{\pgfpoint{1pt}{0pt}}
\pgfsetyvec{\pgfpoint{0pt}{1pt}}
\pgfsetroundjoin 
\pgfsetroundcap
\pgftranslateto{\pgfxy(0,148)}
\begin{pgfmagnify}{1}{-1}
\definecolor{layer0}{rgb}{0.0,0.0,0.0}
\definecolor{layer1}{rgb}{0.0,0.0,0.5}
\definecolor{layer2}{rgb}{1.0,0.0,0.0}
\definecolor{layer3}{rgb}{0.0,0.5,0.5}
\definecolor{layer4}{rgb}{1.0,0.78,0.0}
\definecolor{layer5}{rgb}{0.5,1.0,0.0}
\definecolor{layer6}{rgb}{0.0,1.0,1.0}
\definecolor{layer7}{rgb}{0.0,0.5,0.0}
\definecolor{layer8}{rgb}{0.6,0.8,0.2}
\definecolor{layer9}{rgb}{1.0,0.08,0.58}
\definecolor{layer10}{rgb}{0.71,0.61,0.05}
\definecolor{layer11}{rgb}{0.0,0.5,1.0}
\definecolor{layer12}{rgb}{0.88,0.88,0.88}
\definecolor{layer13}{rgb}{0.64,0.64,0.64}
\definecolor{layer14}{rgb}{0.37,0.37,0.37}
\definecolor{layer15}{rgb}{0.0,0.0,0.0}
\color{layer0}
\pgfsetlinewidth{0.7pt}
\pgfmoveto{\pgfxy(11,23)} 
\pgfcurveto{\pgfxy(14,32)}{\pgfxy(17,41)}{\pgfxy(29,52)}
\pgfstroke
\pgfmoveto{\pgfxy(29.0,52.0)}
\pgflineto{\pgfxy(24.699934182162345,50.7714097663321)}
\pgflineto{\pgfxy(27.402832696231727,47.82279320552914)}
\pgfclosepath 
\pgffill 
\pgfmoveto{\pgfxy(29,52)} 
\pgfcurveto{\pgfxy(36,59)}{\pgfxy(50,63)}{\pgfxy(53,63)}
\pgfstroke
\pgfmoveto{\pgfxy(97,21)} 
\pgfcurveto{\pgfxy(92,32)}{\pgfxy(90,41)}{\pgfxy(77,52)}
\pgfstroke
\pgfmoveto{\pgfxy(77.0,52.0)}
\pgflineto{\pgfxy(78.76166065854412,47.88945846339708)}
\pgflineto{\pgfxy(81.34542962440881,50.94300360487354)}
\pgfclosepath 
\pgffill 
\pgfmoveto{\pgfxy(77,52)} 
\pgfcurveto{\pgfxy(70,59)}{\pgfxy(56,63)}{\pgfxy(53,63)}
\pgfstroke
\pgfline{\pgfxy(53.0,63.0)}{\pgfxy(53.0,84.0)}
\pgfline{\pgfxy(53.0,84.0)}{\pgfxy(53.0,104.0)}
\pgfmoveto{\pgfxy(53.0,84.0)}
\pgflineto{\pgfxy(55.0,88.0)}
\pgflineto{\pgfxy(51.0,88.0)}
\pgfclosepath 
\pgffill 
\begin{pgfmagnify}{1}{-1}
\pgfputat{\pgfxy(29,-41)}{\pgfbox[left,top]{$e_1$}}
\end{pgfmagnify}
\begin{pgfmagnify}{1}{-1}
\pgfputat{\pgfxy(84,-49)}{\pgfbox[left,top]{$e_2$}}
\end{pgfmagnify}
\begin{pgfmagnify}{1}{-1}
\pgfputat{\pgfxy(57,-80)}{\pgfbox[left,top]{$e_3$}}
\end{pgfmagnify}
\begin{pgfmagnify}{1}{-1}
\pgfputat{\pgfxy(174,-60)}{\pgfbox[left,top]{$e_{13}$}}
\end{pgfmagnify}
\begin{pgfmagnify}{1}{-1}
\pgfputat{\pgfxy(211,-6)}{\pgfbox[left,top]{$e_{12}$}}
\end{pgfmagnify}
\begin{pgfmagnify}{1}{-1}
\pgfputat{\pgfxy(252,-60)}{\pgfbox[left,top]{$e_{23}$}}
\end{pgfmagnify}
\pgfline{\pgfxy(176.0,24.0)}{\pgfxy(193.0,62.0)}
\pgfmoveto{\pgfxy(193.0,62.0)}
\pgflineto{\pgfxy(189.54089925597714,59.16545911253682)}
\pgflineto{\pgfxy(193.1921722635568,57.53199487230381)}
\pgfclosepath 
\pgffill 
\pgfline{\pgfxy(193.0,62.0)}{\pgfxy(214.0,105.0)}
\pgfline{\pgfxy(242.0,62.0)}{\pgfxy(221.0,105.0)}
\pgfline{\pgfxy(259.0,24.0)}{\pgfxy(242.0,62.0)}
\pgfmoveto{\pgfxy(242.0,62.0)}
\pgflineto{\pgfxy(241.8078277364432,57.53199487230381)}
\pgflineto{\pgfxy(245.45910074402286,59.16545911253682)}
\pgfclosepath 
\pgffill 
\pgfline{\pgfxy(178.0,20.0)}{\pgfxy(218.0,20.0)}
\pgfmoveto{\pgfxy(218.0,20.0)}
\pgflineto{\pgfxy(214.0,22.0)}
\pgflineto{\pgfxy(214.0,18.0)}
\pgfclosepath 
\pgffill 
\pgfline{\pgfxy(218.0,20.0)}{\pgfxy(259.0,20.0)}
\begin{pgfmagnify}{1}{-1}
\pgfputat{\pgfxy(50,-139)}{\pgfbox[left,top]{$\text{a})$}}
\end{pgfmagnify}
\begin{pgfmagnify}{1}{-1}
\pgfputat{\pgfxy(218,-137)}{\pgfbox[left,top]{$\text{b})$}}
\end{pgfmagnify}
\color{layer2}
\pgfmoveto{\pgfxy(214,104)}
\pgflineto{\pgfxy(223,104)}
\pgflineto{\pgfxy(223,112)}
\pgflineto{\pgfxy(214,112)}
\pgfclosepath 
\pgffill 
\begin{pgfmagnify}{1}{-1}
\pgfputat{\pgfxy(94,-6)}{\pgfbox[left,top]{$2$}}
\end{pgfmagnify}
\pgfmoveto{\pgfxy(92,16)}
\pgflineto{\pgfxy(101,16)}
\pgflineto{\pgfxy(101,24)}
\pgflineto{\pgfxy(92,24)}
\pgfclosepath 
\pgffill 
\begin{pgfmagnify}{1}{-1}
\pgfputat{\pgfxy(172,-6)}{\pgfbox[left,top]{$1$}}
\end{pgfmagnify}
\begin{pgfmagnify}{1}{-1}
\pgfputat{\pgfxy(260,-6)}{\pgfbox[left,top]{$2$}}
\end{pgfmagnify}
\pgfmoveto{\pgfxy(256,16)}
\pgflineto{\pgfxy(265,16)}
\pgflineto{\pgfxy(265,24)}
\pgflineto{\pgfxy(256,24)}
\pgfclosepath 
\pgffill 
\pgfmoveto{\pgfxy(49,102)}
\pgflineto{\pgfxy(57,102)}
\pgflineto{\pgfxy(57,111)}
\pgflineto{\pgfxy(49,111)}
\pgfclosepath 
\pgffill 
\begin{pgfmagnify}{1}{-1}
\pgfputat{\pgfxy(217,-116)}{\pgfbox[left,top]{$3$}}
\end{pgfmagnify}
\pgfmoveto{\pgfxy(171,16)}
\pgflineto{\pgfxy(179,16)}
\pgflineto{\pgfxy(179,24)}
\pgflineto{\pgfxy(171,24)}
\pgfclosepath 
\pgffill 
\pgfmoveto{\pgfxy(7,16)}
\pgflineto{\pgfxy(15,16)}
\pgflineto{\pgfxy(15,24)}
\pgflineto{\pgfxy(7,24)}
\pgfclosepath 
\pgffill 
\begin{pgfmagnify}{1}{-1}
\pgfputat{\pgfxy(50,-114)}{\pgfbox[left,top]{$3$}}
\end{pgfmagnify}
\begin{pgfmagnify}{1}{-1}
\pgfputat{\pgfxy(6,-6)}{\pgfbox[left,top]{$1$}}
\end{pgfmagnify}
\color{layer11}
\pgfellipse[fillstroke]{\pgfxy(53.0,63.0)}{\pgfxy(4.0,0)}{\pgfxy(0,4.0)}
\begin{pgfmagnify}{1}{-1}
\pgfputat{\pgfxy(48,-48)}{\pgfbox[left,top]{$4$}}
\end{pgfmagnify}
\end{pgfmagnify}
\end{pgfpicture}
	\caption{Graphs of a bus-connected ImG with 3 DGUs (on the left) and its equivalent load-connected network (on the right). Red squares indicate DGUs with corresponding local loads $I_{Li}$ (boundary nodes), while the blue circle denotes the unique load at the common bus (internal node).}
	\label{fig:Y_Delta}
\end{figure}

The fact that the method in \cite{riverso2015plug} allows to design stabilizing PnP controllers for load-connected ImGs only, does not represent a limitation. In fact, as shown in \cite{dorfler2013kron,TucciFloriduz_Kron}, arbitrary interconnections of loads and DGUs can always be mapped into their equivalent load-connected topologies by means of Kron reduction \cite{kron1939tensor}. Kron reduction is a well-known method for simplifying linear electrical networks  while preserving the behavior of electrical variables at target nodes (also called \textit{boundary nodes}). In particular, Kron reduction provides an algebraic procedure for computing (i) the corresponding load-connected topology (where only boundary nodes are connected) of the original, arbitrarily connected network, (ii) the admittances of the new edges and (iii) the equivalent currents injected at boundary nodes accounting for the effect of the currents of the eliminated nodes (also called \textit{internal nodes}) in the original network. An example of the transformation is provided in Figure \ref{fig:Y_Delta}.  
\normalsize

 Since several works from the literature consider ImGs arranged in bus-connected topology \cite{7781662,7121016,Vasquez2013}, we are interested in mapping these topologies into the corresponding load-connected ones. While it goes without saying that Kron reduction can also be applied to this aim (by eliminating the node representing the load bus\footnote{Node 4 in the example in Figure \ref{fig:Y_Delta}.}), it requires the inversion of an admittance matrix. In the following, instead, we provide an equivalent procedure, leading to explicit formulae, for computing the admittances of the power lines of the corresponding load-connected network (e.g. the weights of edges $e_{12}, e_{13}$ and $e_{23}$ in Figure \ref{fig:Y_Delta}). This method relies on Kirchhoff's Current Law (KCL), Kirchhoff's Voltage Law (KVL) and QSL approximation.
 
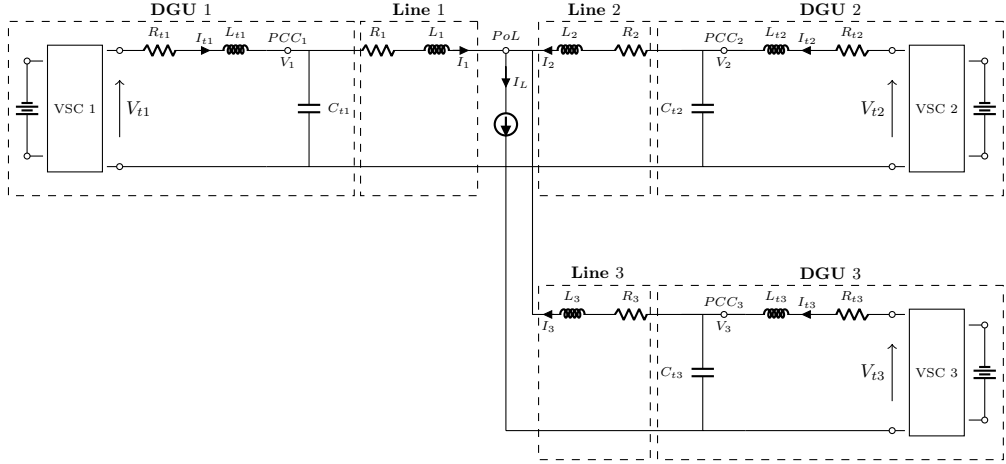
\begin{figure}[!htb]
	\centering 
	\ctikzset{bipoles/length=0.7cm}
\begin{circuitikz}[scale=.7	,transform shape]
\ctikzset{current/distance=1}
\draw

node[] (Ti) at (0,0) {}
node[] (Tj) at ($(9,0)$) {}
node[] (Tk) at ($(9,-5)$) {}
node[] (Tx) at ($(0,-5)$) {}

node[ocirc] (Aibattery) at ([xshift=-4.5cm,yshift=0.9cm]Ti) {}
node[ocirc] (Bibattery) at ([xshift=-4.5cm,yshift=-0.9cm]Ti) {}

(Bibattery) to [battery] (Aibattery) {}
node [rectangle,draw,minimum width=0.8cm,minimum height=2.4cm] (bucki) at ($0.5*(Aibattery)+0.5*(Bibattery)+(0.9,0)$) {\scriptsize{VSC $1$}}
(Aibattery) to [short] ([xshift=0.3cm]Aibattery)
(Bibattery) to [short] ([xshift=0.3cm]Bibattery) 

node[ocirc] (Ai) at ($(Aibattery)+(1.74,0.2)$) {}
node[ocirc] (Bi) at ($(Bibattery)+(1.74,-0.2)$) {}
(Ai) to [short] ([xshift=-0.24cm]Ai)
(Bi) to [short] ([xshift=-0.24cm]Bi)
(Ai) to [R, l=\scriptsize{$R_{t1}$}] ($(Ai)+(1.5,0)$) {}
to [short,i=\scriptsize{$I_{t1}$}]($(Ai)+(1.6,0)$){}
to [L, l=\scriptsize{$L_{t1}$}]($(Ti)+(0,1.1)$){}
(Bi) to [short] ($(Ti)+(0,-1.1)$);
\begin{scope}[shorten >= 10pt,shorten <= 10pt,]
\draw[<-] (Ai) -- node[right] {$V_{t1}$} (Bi);
\end{scope};

\draw
($(Ti)+(0.4,1.1)$) node[anchor=north]{\scriptsize{$V_1$}}
($(Ti)+(0.4,1.1)$) node[anchor=south]{\scriptsize{$PCC_1$}}
($(Ti)+(0.4,1.1)$) node[ocirc](PCCi){}
($(Ti)+(0.8,1.1)$) to [C, l=\scriptsize{$C_{t1}$}] ($(Ti)+(0.8,-1.1)$)

($(Ti)+(0,1.1)$) to [short] ($(Ti)+(0.8,1.1)$)
($(Ti)+(0.8,1.1)$)--($(Ti)+(1.6,1.1)$) to [R, l=\scriptsize{$R_{1}$}] ($(Ti)+(2.6,1.1)$) {}
to [L, l=\scriptsize{$L_{1}$}]($(Tj)+(-5.2,1.1)$){}
($(Tj)+(-5.2,1.1)$) to [short] ($(Tj)+(-5.5,1.1)$)
($(Tj)+(-5.5,1.1)$) to [short,i_=\scriptsize{$I_{1}$}] ($(Tj)+(-5.3,1.1)$)--($(Tj)+(-4.5,1.1)$)

($(Ti)+(4.5,1.1)$)--($(Ti)+(4.5,0.8)$) to [short,i>=\scriptsize{$I_{L}$}]($(Ti)+(4.5,0.5)$)
to [I]($(Ti)+(4.5,-1.1)$)
($(Ti)+(4.5,1.6)$) node[anchor=north]{\scriptsize{$PoL$}}

node[ocirc] at ($(Ti)+(4.5,1.1)$) {}

($(Ti)+(4.5,1.1)$)--($(Ti)+(5,1.1)$) to [L, l=\scriptsize{$L_{2}$}]($(Ti)+(6.4,1.1)$) {}
to [R, l=\scriptsize{$R_{2}$}] ($(Ti)+(7.3,1.1)$) 
($(Tj)+(-1.2,1.1)$) to [short] ($(Tj)+(0,1.1)$)
($(Ti)+(0,-1.1)$) to [short] ($(Tj)+(0,-1.1)$)
($(Tj)+(0,1.1)$) to [short] ($(Tj)+(-1.2,1.1)$)
($(Tj)+(-1.2,1.1)$) to [short] ($(Tj)+(-1.7,1.1)$)
($(Tj)+(-3.6,1.1)$) to [short,i=\scriptsize{$I_{2}$}] ($(Tj)+(-3.7,1.1)$)

($(Tj)+(-0.4,1.1)$) node[anchor=north]{\scriptsize{$V_2$}}
($(Tj)+(-0.4,1.1)$) node[anchor=south]{\scriptsize{$PCC_2$}}
($(Tj)+(-0.4,1.1)$) node[ocirc](PCCj){}
($(Tj)+(-0.8,1.1)$) to [C, l_=\scriptsize{$C_{t2}$}] ($(Tj)+(-0.8,-1.1)$)

node[ocirc] (Ajbattery) at ([xshift=4.5cm,yshift=0.9cm]Tj) {}
node[ocirc] (Bjbattery) at ([xshift=4.5cm,yshift=-0.9cm]Tj) {}
(Bjbattery) to [battery] (Ajbattery) {}
node [rectangle,draw,minimum width=0.8cm,minimum height=2.4cm] (vsci) at ($0.5*(Ajbattery)+0.5*(Bjbattery)-(0.9,0)$) {\scriptsize{VSC $2$}}
(Ajbattery) to [short] ([xshift=-0.3cm]Ajbattery)
(Bjbattery) to [short] ([xshift=-0.3cm]Bjbattery)

node[ocirc] (Aj) at ($(Ajbattery)+(-1.74,0.2)$) {}
node[ocirc] (Bj) at ($(Bjbattery)+(-1.74,-0.2)$) {}
(Aj) to [short] ([xshift=0.24cm]Aj)
(Bj) to [short] ([xshift=0.24cm]Bj)
(Bj) to [short] ($(Tj)+(0,-1.1)$)
(Aj) to [R, l_=\scriptsize{$R_{t2}$}] ($(Aj)+(-1.5,0)$) {}
to [short,i_=\scriptsize{$I_{t2}$}]($(Aj)+(-1.6,0)$){}
($(Tj)+(0,1.1)$) to [L, l=\scriptsize{$L_{t2}$}]($(Aj)+(-1.6,0)$){};
\begin{scope}[shorten >= 10pt,shorten <= 10pt,]
\draw[<-] (Aj) -- node[left] {$V_{t2}$} (Bj);
\end{scope};

\draw
node [rectangle,draw,minimum width=6.5cm,minimum height=3.3cm,dashed,label=\small\textbf{DGU $1$}] (DGUi) at ($0.5*(Aibattery)+0.5*(Bibattery)+(2.9,0)$) {}
node [rectangle,draw,minimum width=6.5cm,minimum height=3.3cm,dashed,label=\small\textbf{DGU $2$}] (DGUj) at ($0.5*(Ajbattery)+0.5*(Bjbattery)-(2.9,0)$) {}
node [rectangle,draw,minimum width=2.2cm,minimum height=3.3cm,dashed,label=\small\textbf{Line $1$ }] (Lineij) at ($1.52*(DGUi.center)+0.5*(DGUj.center)+(0,0)$){}
node [rectangle,draw,minimum width=2.1cm,minimum height=3.3cm,dashed,label=\small\textbf{Line $2$ }] (Lineij) at ($-0.54*(DGUi.center)+0.5*(DGUj.center)+(0,0)$){}

node[ocirc] (Akbattery) at ([xshift=4.5cm,yshift=0.9cm]Tk) {}
node[ocirc] (Bkbattery) at ([xshift=4.5cm,yshift=-0.9cm]Tk) {}
(Bkbattery) to [battery] (Akbattery) {}
node [rectangle,draw,minimum width=0.8,minimum height=2.4cm] (vsci) at ($0.5*(Akbattery)+0.5*(Bkbattery)-(0.9,0)$) {\scriptsize{VSC $3$}}
(Akbattery) to [short] ([xshift=-0.3cm]Akbattery)
(Bkbattery) to [short] ([xshift=-0.3cm]Bkbattery)

node[ocirc] (Ak) at ($(Akbattery)+(-1.74,0.2)$) {}
node[ocirc] (Bk) at ($(Bkbattery)+(-1.74,-0.2)$) {}
(Ak) to [short] ([xshift=0.24cm]Ak)
(Bk) to [short] ([xshift=0.24cm]Bk)
(Bk) to [short] ($(Tk)+(0,-1.1)$)
(Ak) to [R, l_=\scriptsize{$R_{t3}$}] ($(Ak)+(-1.5,0)$) {}
to [short,i_=\scriptsize{$I_{t3}$}]($(Ak)+(-1.6,0)$){}
($(Tk)+(0,1.1)$) to [L, l=\scriptsize{$L_{t3}$}]($(Ak)+(-1.6,0)$){};
\begin{scope}[shorten >= 10pt,shorten <= 10pt,]
\draw[<-] (Ak) -- node[left] {$V_{t3}$} (Bk);
\end{scope};

\draw

($(Tx)+(5,1.1)$)--($(Tx)+(5.1,1.1)$) to [L, l=\scriptsize{$L_{3}$}]($(Tx)+(6.4,1.1)$) {}
to [R, l=\scriptsize{$R_{3}$}] ($(Tx)+(7.3,1.1)$) 

($(Tk)+(-1.2,1.1)$) to [short] ($(Tk)+(0,1.1)$)
($(Tx)+(4.5,-1.1)$) to [short] ($(Tk)+(0,-1.1)$)
($(Tk)+(0,1.1)$) to [short] ($(Tk)+(-1.2,1.1)$)
($(Tk)+(-1.2,1.1)$) to [short] ($(Tk)+(-1.7,1.1)$)
($(Tk)+(-3.6,1.1)$) to [short,i=\scriptsize{$I_{3}$}] ($(Tk)+(-3.7,1.1)$)
($(Ti)+(4.5,-1.1)$) to [short] ($(Tk)+(-4.5,-1.1)$)
($(Tx)+(5,1.1)$) to [short] ($(Ti)+(5,1.1)$)

($(Tk)+(-0.4,1.1)$) node[anchor=north]{\scriptsize{$V_3$}}
($(Tk)+(-0.4,1.1)$) node[anchor=south]{\scriptsize{$PCC_3$}}
($(Tk)+(-0.4,1.1)$) node[ocirc](PCCk){}
($(Tk)+(-0.8,1.1)$) to [C, l_=\scriptsize{$C_{t3}$}] ($(Tk)+(-0.8,-1.1)$)

node [rectangle,draw,minimum width=6.5cm,minimum height=3.3cm,dashed,label=\small\textbf{DGU $3$}] (DGUk) at ($0.5*(Akbattery)+0.5*(Bkbattery)-(2.9,0)$) {}
node [rectangle,draw,minimum width=2.1cm,minimum height=3.3cm,dashed,label=\small\textbf{ Line $3$ }] (Lineij) at ($-0.54*(DGUi.center)+0.5*(DGUj.center)+(0,-5)$){}

;\end{circuitikz} 
	\caption{Electrical scheme of a bus-connected ImG composed of three DGUs and a common unmodeled load.}
	\label{fig:3DGUbusconnected}
\end{figure}

	As a starting point, let us consider a bus-connected ImG with $N$ DGUs feeding a common load ($I_{L}$) connected to the Point of Load (PoL). Figure \ref{fig:3DGUbusconnected} provides an example with $N=3$ and shows parameters $R_i$, $L_i$ and currents $I_i$ characterizing each DGU. As usual, we assume balanced lines. By applying KCL and KVL in the $abc$-frame and performing the Park's transformation \cite{Park1929}, we have that the dynamics of DGU $i$ in $dq$-coordinates, with $\omega=\omega_0$, are described by
\begin{equation}
\label{eq:BC_model1} \frac{\mathrm d}{\mathrm d t}V_{i}^{dq} = \frac{I_{ti}^{dq}}{C_{ti} }-\frac{I_{ i}^{dq}}{C_{ti}} -  \mathrm i \omega_0 V_i^{dq}
\end{equation}
and \eqref{eq:LC_model_2}. For $1\leq j \leq N$, $j \neq i$, it holds
\begin{equation}
\label{eq:line_currents_BC}
V_i^{dq}-V_j^{dq}=
R_{ i}I_{ i}^{dq}+
L_{ i} \frac{\mathrm d}{\mathrm d t}I_{ i}^{dq} +
\mathrm i \omega_0 L_{ i} I_{ i}^{dq}
-
R_{ j}I_{j}^{dq}-
L_{ j} \frac{\mathrm d}{\mathrm d t}I_{ j}^{dq} -
\mathrm i \omega_0 L_{ j} I_{ j}^{dq}. 
\end{equation}
By virtue of QSL approximations (see \cite{riverso2015plug,Venkatasubramanian1995}), in \eqref{eq:line_currents_BC}, we set $\frac{\mathrm d}{\mathrm d t}I_{i}^{dq}=0$,  $\forall i\in \mathcal{V}_{DGU}$. Hence, equation \eqref{eq:line_currents_BC} becomes
\begin{equation*}
	\centering
	V_i^{dq}-V_j^{dq}=
	R_{ i}I_{ i}^{dq}+
	\mathrm i \omega_0 L_{ i} I_{ i}^{dq}-
	R_{ j}I_{ j}^{dq}-
	\mathrm i \omega_0 L_{ j} I_{ j}^{dq}
	\;,\qquad 1\leq j \leq N, \; j \neq i.
\end{equation*}
Next, defining $Z_i=R_{i}+\mathrm i\, \omega_0 L_{i}$, one gets
\begin{equation}
I_j^{dq}=\frac{V_j^{dq}-V_i^{dq}}{Z_j}+
\frac{Z_i}{Z_j}I_i^{dq}
\;,\qquad 1\leq j \leq N, \; j \neq i.
\label{chap3.3:Ij}
\end{equation}
Applying KCL at the PoL, we have
\begin{equation}
I_i^{dq}+\sum_{j\neq i} I_j^{dq}=I_L^{dq}
\label{chap3.3:KCLiload}
\end{equation}
and, inserting \eqref{chap3.3:Ij} in \eqref{chap3.3:KCLiload}, we obtain
\begin{equation}
\label{eq:I_i}
	I_i^{dq}=\frac{I_L^{dq}}{\left(1+ \sum_{j\neq i} \frac{Z_i}{Z_j} \right) }+ 
  \left(\sum_{j\neq i} \frac{1}{Z_j} \right)
	\frac{	V_i^{dq}}{\left( 1+ \sum_{j\neq i} \frac{Z_i}{Z_j} \right)}
	-
	\sum_{j\neq i} \frac{V_j^{dq}}{Z_j} 
	\frac{1}{\left( 1+ \sum_{j\neq i} \frac{Z_i}{Z_j} \right)}.
\end{equation}
Observing that, in the formula above
\begin{equation*}
	\frac{1}{\left( 1+ \sum_{j\neq i} \frac{Z_i}{Z_j} \right)}=
	\frac{1}{\left(Z_i \sum_{k=1}^N \frac{1}{Z_k} \right)},
\end{equation*}
we can introduce the following notation

	\begin{equation*}
	\label{eq:equivalent_line_imp}
	\frac{1}{Z_{ij}}=  
	\frac{1}{Z_j Z_i \sum_{k=1}^N \frac{1}{Z_k}}\; , \forall j \neq i,	
\end{equation*}
which implies
	\begin{equation}
\label{eq:sum_admittances_Yij}
	 \left(\sum_{j\neq i} \frac{1}{Z_j} \right) 
	\frac{1}{\left(Z_i \sum_{k=1}^N \frac{1}{Z_k} \right)}
	=\sum_{j \neq i} \frac{1}{Z_{ij}}.
\end{equation}

At this point, by replacing \eqref{eq:sum_admittances_Yij} in \eqref{eq:I_i}, and then substituting the resulting expression into \eqref{eq:BC_model1}, one gets that the dynamics of $V_i^{dq}$ have the form \eqref{eq:LC_model_1}. In other words, we have derived the equivalent load-connected model of a bus-connected ImG with complex power lines admittances $1/Z_i$, $i=1,...,N$. Notably, the weights of the equivalent load-connected network (indicated with $1/Z_{ij}$ $i,j=1,...,N,\; i\neq j$, as in \eqref{eq:LC_model_1}) are given by \eqref{eq:sum_admittances_Yij}, while the effect of the eliminated node (i.e. the PoL) at the $i$-th PCC is accounted by the following injected current
	\begin{equation}
		\label{eq:BC_model_notations}
		I_{Li}^{dq}= \frac{I_L^{dq}}{\left(Z_i \sum_{k=1}^N \frac{1}{Z_k} \right)}.
	\end{equation}
 Finally, we highlight that, since admittances $1/Z_{ij}$ $i,j=1,...,N,\; i\neq j$ are all nonzero, the described transformation always returns a fully connected reduced network (see the example in Figure \ref{fig:Y_Delta}-b, where each pair of distinct nodes is connected by a unique edge).

\section{Plug-and-play primary control layer}
\label{sec:PnP_primary}
\subsection{Control architecture}
\label{sec:ctrl_architecture}
\begin{figure}[!htb]
	\centering
	\includegraphics[scale=0.5]{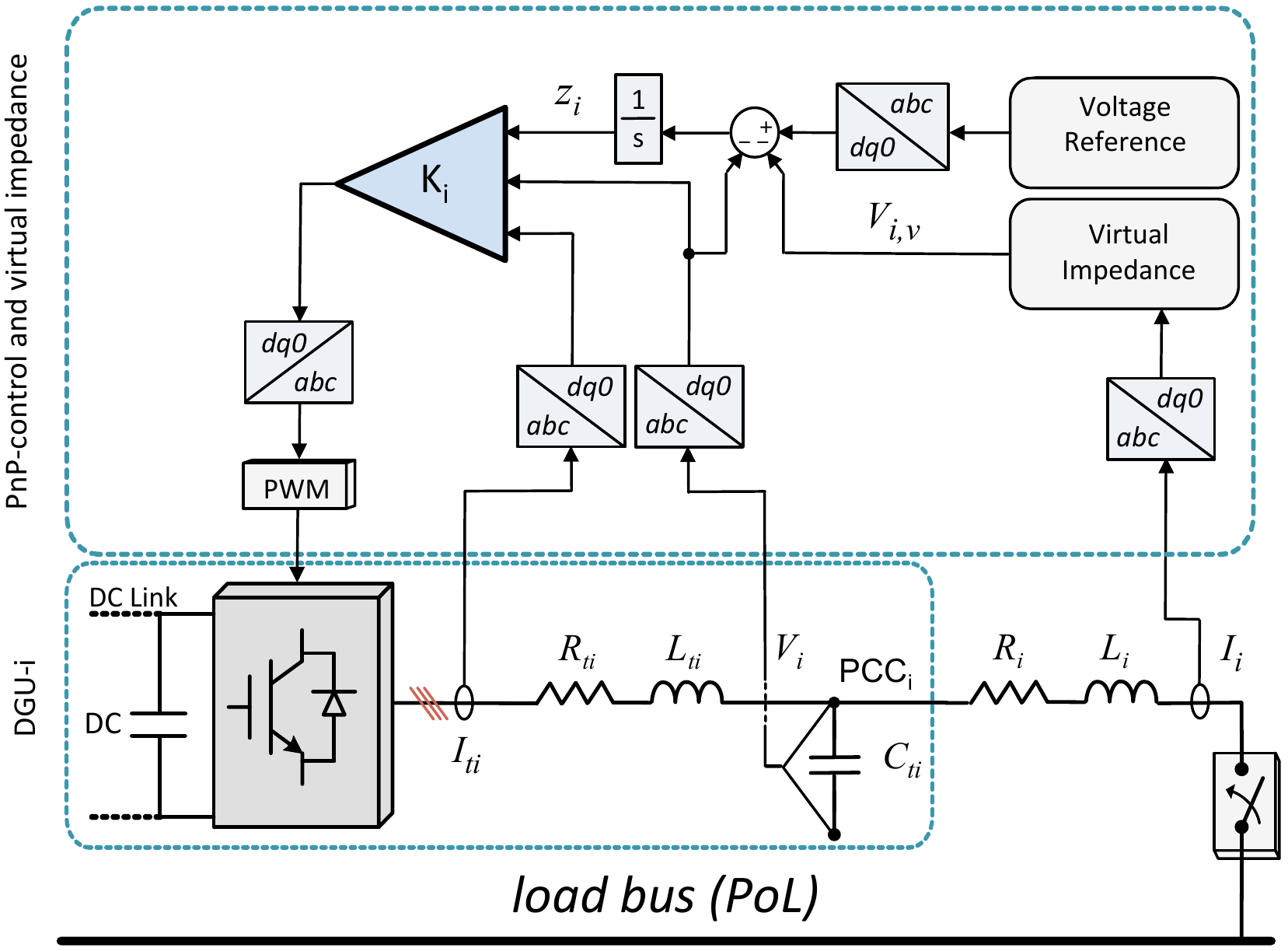}
	\caption{Bus-connected DGU equipped with local PnP regulator and virtual impedance loop.}
	\label{fig:DGUctrl}
\end{figure}

Figure \ref{fig:DGUctrl} shows a bus-connected DGU, equipped with a decentralized PnP controller. Each local regulator exploits measurements of the voltage $V^{dq}_i$ at the PCC and the current $I^{dq}_{ti}$, in order to control the voltage $V_{ti}^{dq}$ at the VSC $i$ and make $V_{i}^{dq}$ track a reference signal. We notice that the proposed controller is multivariable, where the only tunable parameter is the matrix gain $K_i$ in Figure \ref{fig:DGUctrl}. Compared to common solutions found in the literature \cite{Guerrero2013,Vasquez2013}, the proposed scheme breaks the hierarchical structure composed of inner current and voltage loops and droop control. However, as in single-input single-output current and voltage loops, feedback information consists of the current provided by the VSC and the voltage at the PCC. Most importantly, the presence of the integrators guarantees tracking of constant set-point references in the $dq$-frame, and hence of three-phase sinusoidal quantities in the $abc$-frame. 

Figure \ref{fig:DGUctrl} also reveals the presence of a virtual impedance loop. Virtual impedance is a widely used tool in control of parallel interconnected VSCs \cite{He2002,Vasquez2013}. Its employment is instrumental for lowering the circulating currents between DGUs generated by line-impedance unbalance and mismatch in inverter parameters. In practice, as shown in Figure \ref{fig:DGUctrl}, the virtual impedance reduces the voltage reference by a term proportional to the line current, thus mimicking an $RL$ impedance connected in series to the output filter of the inverter. We also highlight that virtual resistances and inductances should be chosen sufficiently large so as to outnumber such uncertainties in the electrical parameters. In contrast with a physical device, a virtual impedance has no power losses.
In \cite{He2002,Vasquez2013}, the virtual impedance is expressed in $\alpha\beta$-coordinates. Similarly, we can model it in $dq$-coordinates as follows:
\begin{equation}
\begin{aligned}
V_{i,v}^d &= R_{i,v} I_{i}^d + L_{i,v}\frac{\mathrm{d}I_{i}^d}{\mathrm{d}t} - \omega_0 L_{i,v} I_{i}^q\\
V_{i,v}^q &= R_{i,v} I_{i}^q + L_{i,v}\frac{\mathrm{d}I_{i}^q}{\mathrm{d}t} + \omega_0 L_{i,v} I_{i}^d
\end{aligned}
\end{equation}
where $R_{i,v}$ and $L_{i,v}$ are the virtual resistance and inductance parameters and $V_{i,v}$ and $I_{i}$ are the voltage and current in the $dq$-frame shown in Figure \ref{fig:DGUctrl}. 
\subsection{Plug-and-play design of local controllers}
\label{sec:PnP_design}
In the following, we first summarize the PnP algorithm in \cite{riverso2015plug} for designing the matrix gain $K_i$ in Figure \ref{fig:DGUctrl}. Then, since the algorithm in \cite{riverso2015plug} assumes load-connected ImGs only, we exploit the result derived in Section \ref{sec:load-connected_to_bus-connected} to apply the above methodology to bus-connected networks.\\ 
PnP design of local controllers proposed in \cite{riverso2015plug} guarantees overall voltage and frequency stability, with the main advantage that a global ImG model is not required in any design step. Moreover, individual DGUs can test if the addition/removal of a subsystem is dangerous for the ImG stability by following the procedure below.

Assume that the plug-in of a DGU (say DGU $i$) is required. This new unit issues a plug-in request to its future neighboring DGUs $j$, $j\in\mathcal{N}_i$, and then solves the Linear Matrix Inequality (LMI) test (19) in \cite{riverso2015plug}, which is a convex optimization problem depending only upon the parameters of power lines $ij$, $j\in\mathcal{N}_i$ (e.g., $R_{ij}$ and $L_{ij}$ in Figure \ref{fig:load_connected_model}). If the optimization is feasible, the above test returns the matrix gain $K_i$. Moreover, since also DGUs $j\in\mathcal{N}_i$ will have a new neighbor, they must retune their controller, so as to account for the presence of new coupling terms. The update of the $j$-th controller is done by solving an LMI problem analogous to the one solved for DGU $i$. If one of the above LMI tests is unfeasible, the plug-in of DGU $i$ is denied. Otherwise, DGU $i$ (equipped with controller $K_i$) can be connected to the network without compromising the collective voltage and frequency stability of the ImG. 

Unplugging of a DGU (say DGU $m$) follows a similar procedure. As line $mk$ is disconnected, only DGUs $k$, $k\in\mathcal{N}_m$ must successfully recompute their local controllers before allowing the disconnection of DGU $k$ (see \cite{riverso2015plug} for details).

Let us consider now the case in which we have a bus-connected ImG. Whenever we want to plug-in a new DGU (say DGU $i$), the first step is to compute the line impedances between DGU $i$ and each DGU $j$, $j\in\mathcal{V}_{DGU}\setminus\{i\}$, using \eqref{eq:sum_admittances_Yij}. Subsequently, the LMI test (19) in \cite{riverso2015plug} must be successfully solved for all the DGUs, in order to allow the safe connection of DGU $i$. On the other hand, if only one of these tests LMI is infeasible, the plug-in of DGU $i$ is denied. Unplugging of a DGU can be performed in a similar way.

\begin{rmk}
\label{rmk:hot_plug_in}
The procedure for handling plug-in/-out operations when the original network is arranged either in a load- or bus-connected topology can be simplified as follows. Let us assume that the hot plug-in (i.e. the plug-in in real-time) of DGU $i$ has been allowed and scheduled at a future time $\bar t$. At the same time instant, DGUs $j$ (with $j\in\mathcal N_i$ if the ImG is load-connected, or $j\in\mathcal{V}_{DGU}\setminus\{i\}$ if the ImG is bus-connected) will start using the new gains. Retuning of gains $K_j$, however, could be avoided if previous gains are still feasible for the corresponding LMI test. In other words, for such DGUs $j$, one can check if matrix gains $K_j$ working for $t<\bar t$ still fulfill the constraints of their corresponding optimization problem (19) in \cite{riverso2015plug}. Preserving previously designed controllers, in fact, has the advantage of reducing perturbations on electrical signals right after $\bar t$, which might be caused by controller switching.

In a similar way, if DGU $m$ is disconnected, the retuning of controllers of DGUs $k$ (with $k\in\mathcal{N}_m$ or $k\in\mathcal{V}_{DGU}\setminus\{m\}$ if the ImG is load- or bus- connected, respectively) can be avoided if the LMI test (19) in \cite{riverso2015plug} is feasible for previously designed matrix gains $K_k$.
\end{rmk}

\subsection{Clock synchronization for primary control}
\label{sec:synchronization_process}
The computation of VSC commands assumes that all clocks of local controllers, used for performing $abc$ to $dq$ transformations, are synchronized. As also highlighted in \cite{Etemadi2012a}, each DGU can include a crystal oscillator that generates the angular phase $\theta(t)=\int_0^t \omega_0~d\tau$. However, if the local oscillators are not synchronized, the local angular phase is given by $\theta_i(t)=\theta(t)+\theta_{i,0}$, where $\theta_{i,0}$ is the initial offset. In \cite{Etemadi2012a}, the authors propose a synchronization using GPS radio clock, which could achieve an accuracy higher than $1\mu$s \cite{Phadkeand2008}. 

An alternative solution is to synchronize clocks using communication between controllers. This operation can be performed quite rarely because, as noted in \cite{Etemadi2012a}, currently available crystal oscillators are characterized by high accuracy (from 2 $\mu$s to 20 $p$s in a year \cite{Vig2000}). Moreover, synchronization can be done through packet networks, using either a distributed protocol (e.g. Berkeley algorithm \cite{gusella1989accuracy}) or approaches based on all-to-all communication, such as instantaneous averaging \cite{7835645}.

If an ImG is equipped with primary PnP regulators, prior to allowing the plug-in of a new DGU, it is mandatory to synchronize it with all the ones already connected to the PoL. Therefore, in the experiments described in Section \ref{sec:exp_res}, we let each new DGU estimate $\theta_{i,0}$ by computing the average angular phases of all the other DGUs.

\subsection{Harmonic compensation by tuning the PnP control bandwidth}
\label{subsec:ctrl_experiments}
\begin{figure}[!htb]
	\centering
	\begin{subfigure}[!htb]{0.49\textwidth}
		\centering
		\includegraphics[width=1\textwidth]{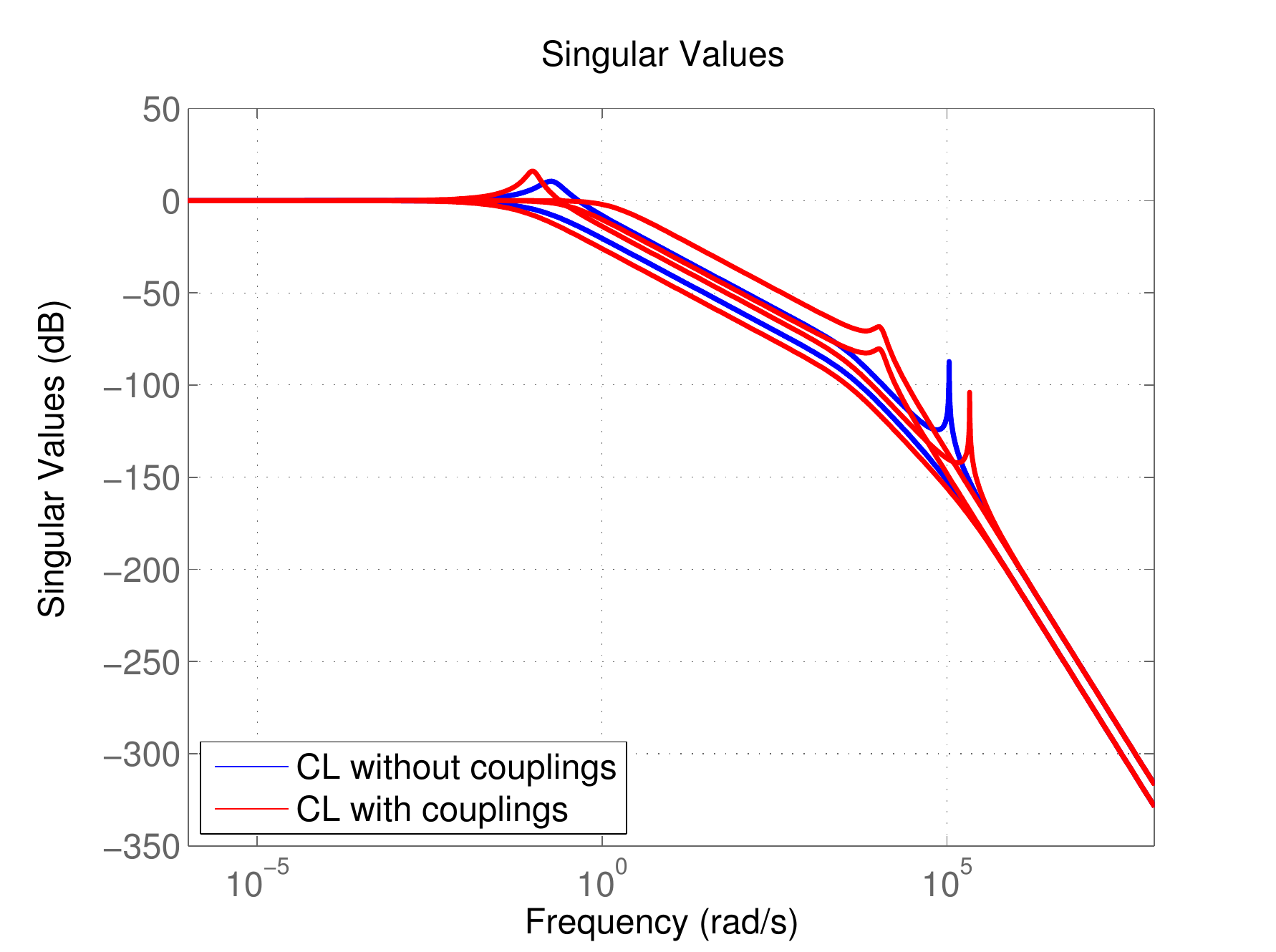}
		\caption{Singular values of the transfer function matrix from $dq$ voltage references to $dq$ voltages at 
			the PCCs.}
		\label{fig:singValVoltage}
	\end{subfigure}
	\begin{subfigure}[!htb]{0.49\textwidth}
		\centering
		\includegraphics[width=1\textwidth]{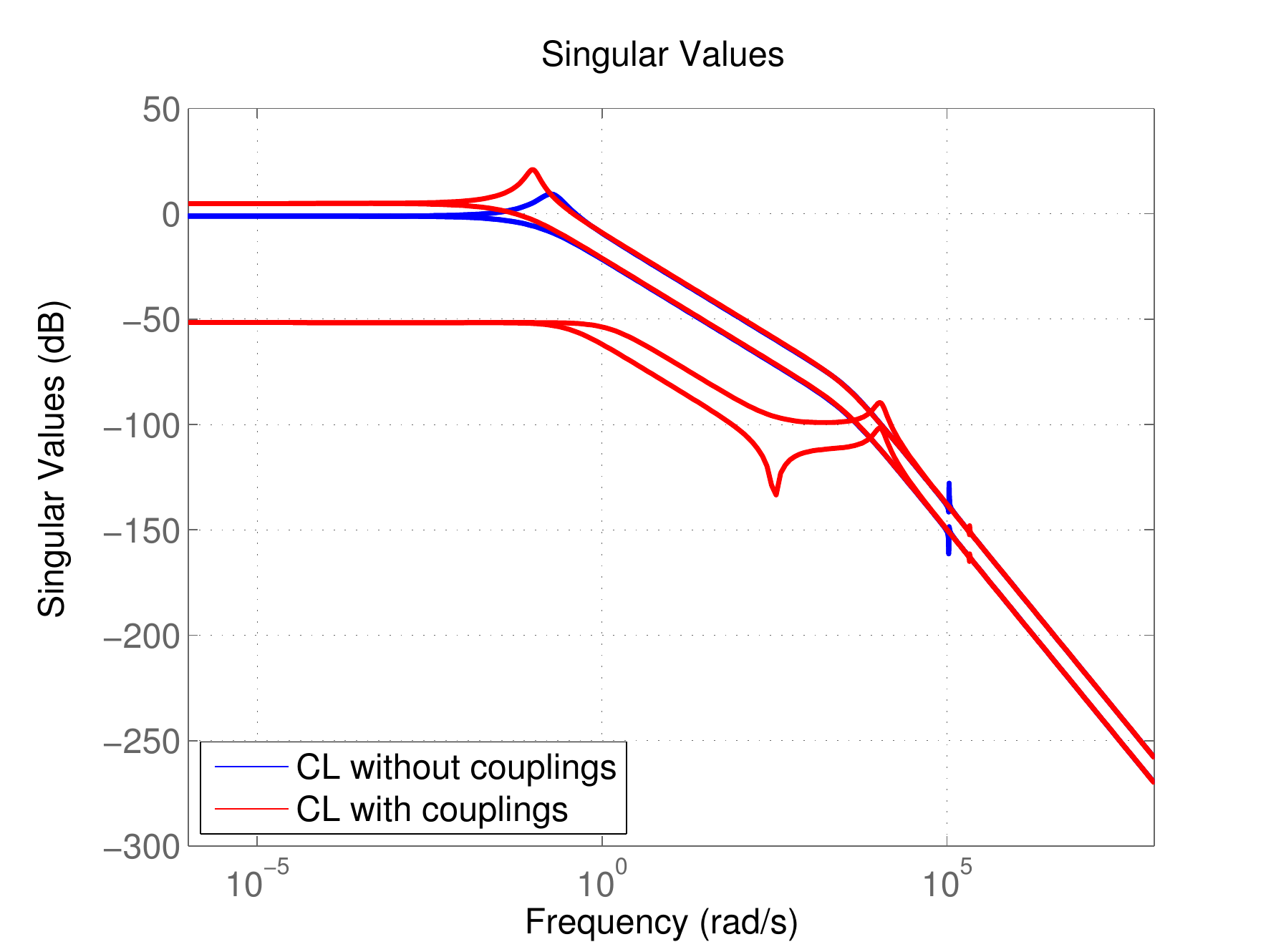}
		\caption{Singular values of the transfer function matrix from $dq$ voltage references to $dq$ currents in the filters.}
		\label{fig:singValCurrent}
	\end{subfigure}
	\caption{Singular values for the closed-loop ImG with two DGUs.}
	\label{fig:singVal}
\end{figure}

Besides collective voltage and frequency stability, PnP design can guarantee good harmonic compensation, even in absence of low-level resonant controllers (described, e.g., in \cite{Vasquez2013}). In fact, as explained in \cite{riverso2015plug}, one can also shape,  in a desired fashion, the singular values of the overall closed-loop ImG. More specifically, in order to provide suitable attenuation of the $5$-th, $7$-th and $11$-th harmonics in $abc$-frame, we design primary PnP controllers that attenuate the $4$-th, $6$-th and $10$-th harmonics in the rotating $dq$-frame. An example is provided in Figure \ref{fig:singVal} for an ImG with two DGUs. The singular values of closed-loop ImG transfer function from voltage references to the voltages at the PCCs are represented in Figure \ref{fig:singValVoltage}. Similarly, the singular values of the transfer function from voltage references to the currents in the filters are given in Figure \ref{fig:singValCurrent}. Notice that good attenuation can be obtained also when couplings between DGUs are accounted for. In fact, assuming $f_0= 50$ Hz, from Figure \ref{fig:singValVoltage}, the attenuation of the 5-th, 7-th and 11-th voltage harmonics is 50 dB, 60 dB and 70 dB, respectively.

\section{Coordinated control}
\label{sec:coordinatedControl}
Using PnP controllers described in Section \ref{sec:PnP_design}, we are able to guarantee voltage and frequency stability for the overall ImG. Notably, each local controller regulates the voltage at the corresponding PCC, according to the following reference in the $abc$-frame
\begin{equation}
\label{eq:PCC_voltage_reference}
\begin{aligned}
V^*_{i_a}(t) &= V_i^*\sin\left(\omega_0t + \phi_i^*\right)\\
V^*_{i_b}(t) &= V_i^*\sin\left(\omega_0t + \phi_i^* - \frac{2\pi}{3} \right)\\
V^*_{i_c}(t) &= V_i^*\sin\left(\omega_0t + \phi_i^* + \frac{2\pi}{3} \right).
\end{aligned}
\end{equation}
In bus-connected ImGs, one can indirectly control voltage and frequency at the PoL by choosing voltages $V^*_i$, $i=1,\ldots,N$ in \eqref{eq:PCC_voltage_reference}. In several control architectures, the use of a Power-Management System (PMS) has been proposed for this purpose (see for example \cite{Etemadi2012a}). The basic idea is to compute voltage references such that each DGU injects prescribed active and reactive power. The PMS must be run in real-time in order to maintain a prescribed power flow, even if loads change. In the following, we propose a distributed secondary control layer capable to guarantee (i) a desired voltage at the PoL, and (ii) sharing of reactive power injections among DGUs.

\subsection{Voltage tracking at the PoL}
\label{sec:restoringPCC}
\begin{figure}[!htb]
	\centering
	\includegraphics[scale=0.42]{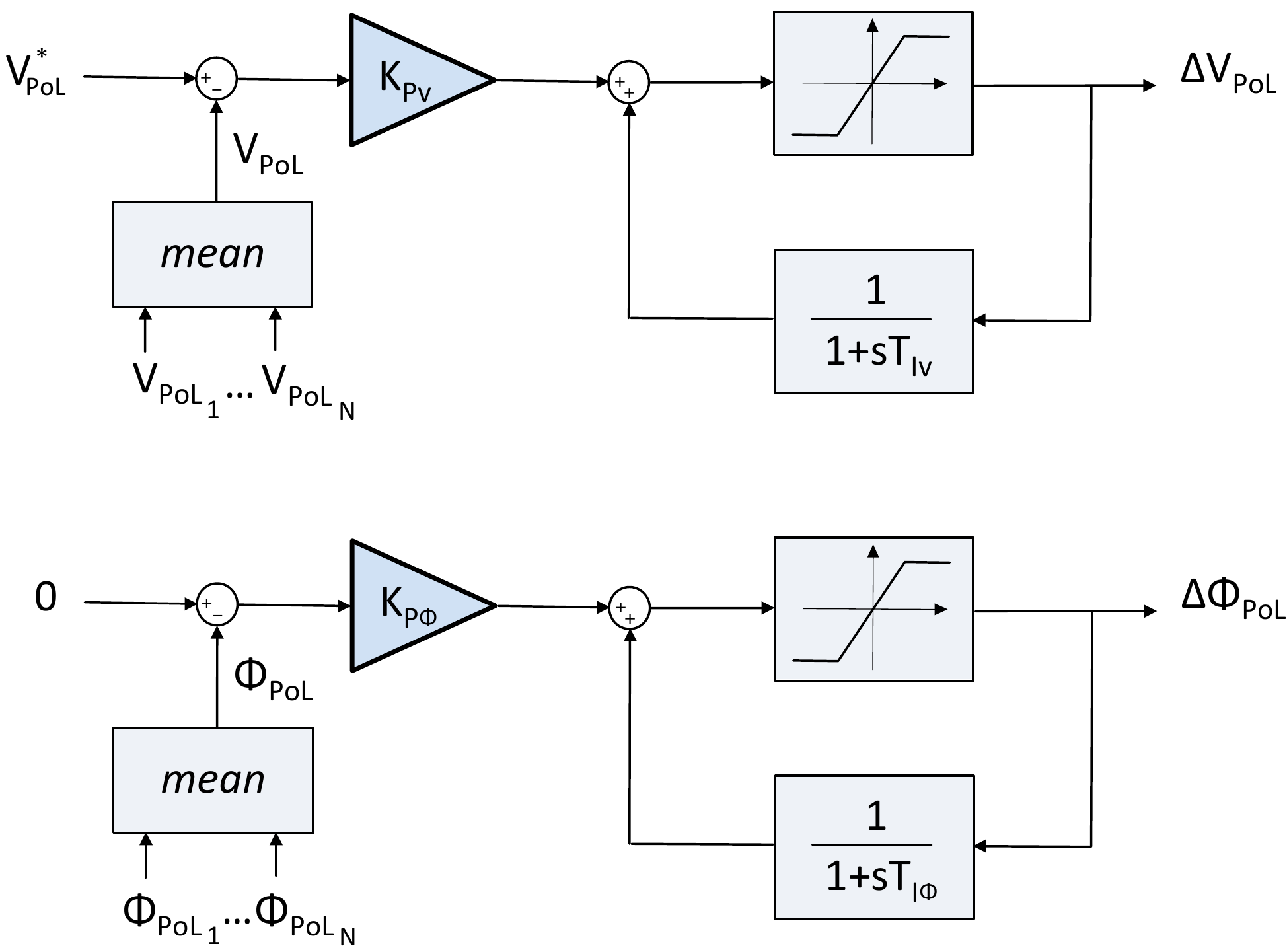}
	\caption{Coordinated control layer: computation of correction terms $\Delta V_{PoL}$ and $\Delta \phi_{PoL}$. Parameters $K_{PV}$ and $K_{P\phi}$ are the voltage and phase proportional coefficients, while $T_{IV}$ and $T_{I\phi}$ are the voltage and phase integral time constants.}
	\label{fig:v_restoring}
\end{figure}

\begin{figure}[!htb]
	\centering
	\includegraphics[scale=0.3]{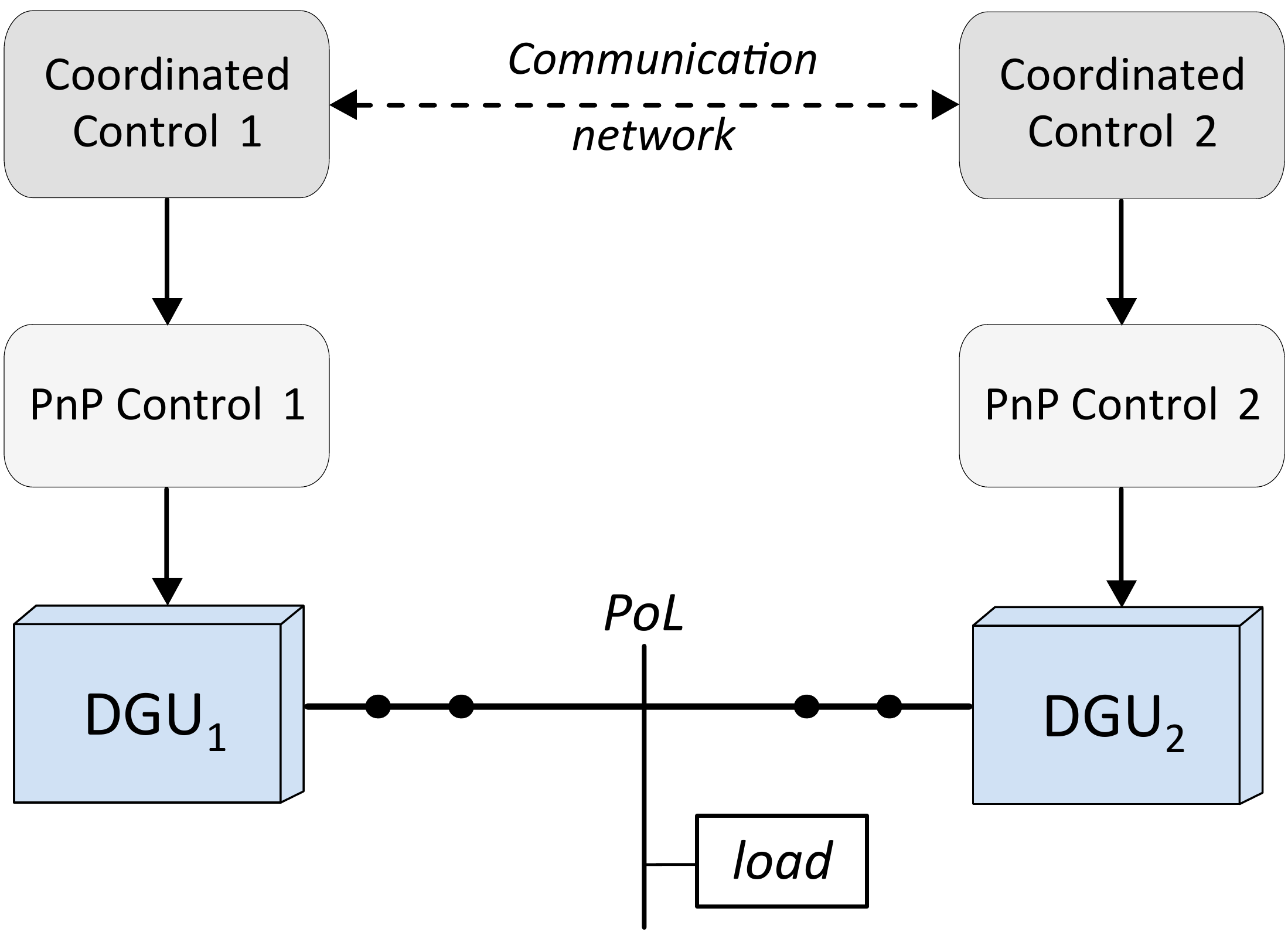}
	\caption{Scheme of the coordinated control.}
	\label{fig:overall_architecture}
\end{figure}

Let us indicate the desired PoL voltage with $V_{PoL}^*\angle{0}$\footnote{Without loss of generality, the phase has been assumed equal to zero.}. In absence of loads, in order to guarantee the reference at the PoL, we could set, in \eqref{eq:PCC_voltage_reference}, $V_i^*=V_{PoL}^*$ and $\phi_{i}^*=0$. However, due to the presence of time-varying loads, $V_i^*$ and $\phi_{i}^*$ must be adapted over time. We propose that each DGU changes its set-point according to
\begin{equation}
\label{eq:voltage_ref_coordinated_control}
\begin{aligned}
V^*_{i_a}(t) &= \left(V_{PoL}^*+\Delta V_{PoL}\right)\sin\left(\omega_0t + \Delta\phi_{PoL}\right)\\
V^*_{i_b}(t) &= \left(V_{PoL}^*+\Delta V_{PoL}\right)\sin\left(\omega_0t + \Delta\phi_{PoL} - \frac{2\pi}{3}\right )\\
V^*_{i_c}(t) &= \left(V_{PoL}^*+\Delta V_{PoL}\right)\sin\left(\omega_0t + \Delta\phi_{PoL} + \frac{2\pi}{3} \right),
\end{aligned}
\end{equation}	
instead of using \eqref{eq:PCC_voltage_reference}. The next aim is to compute $\Delta V_{PoL}$ and $\Delta\phi_{PoL}$ in order to keep PoL voltage close to its reference. Since we can not measure the voltage at the PoL, we estimate its amplitude and phase averaging local measurements
\begin{equation}
\label{eq:averagesPCC}
V_{PoL} = \sum_{i=1}^N \frac{V_{PoL,i}}{N}\qquad \phi_{PoL} = \sum_{i=1}^N \frac{\phi_{PoL,i}}{N}
\end{equation}
where $V_{PoL,i}$ and $\phi_{PoL,i}$ are computed by each DGU from $V_{i}$ and $I_{i}$ (shown in Figure \ref{fig:DGUctrl}) as follows
\begin{equation}
\label{eq:pcci}
V_{PoL,i} = \sqrt{(V^d_{PoL,i})^2+(V^q_{PoL,i})^2}\qquad \phi_{PoL,i}=\frac{V^q_{PoL,i}}{V^d_{PoL,i}},
\end{equation}
where
\begin{equation*}
\begin{aligned}
V^d_{PoL,i} = V^d_{i} + \omega_0L_{i}I^q_{i}\\
V^q_{PoL,i} = V^q_{i} - \omega_0L_{i}I^d_{i}.
\end{aligned}
\end{equation*}
We equip each DGU with the local controller in Figure \ref{fig:v_restoring} for computing $\Delta V_{PoL}$ and $\Delta\phi_{PoL}$ in \eqref{eq:voltage_ref_coordinated_control}. We note that the controller in Figure \ref{fig:v_restoring} is replicated in each DGU instead of being unique for the whole ImG. As shown in \cite{Shafiee2014}, replicating the controller has several advantages when the communication latency increases.
Differently from the PnP control architecture that is completely decentralized, the secondary layer of controllers is distributed as it needs a communication network in order to exchange values $V_{PoL,i}$ and $\phi_{PoL,i}$, and then to compute locally the averages \eqref{eq:averagesPCC}. Formula \eqref{eq:averagesPCC} requires a fully connected communication network, as all measurements $V_{PoL,i}$ and $\phi_{PoL,i}$ have to be broadcasted to all DGUs. However, this limitation could be avoided by resorting to distributed algorithms, based on consensus strategies, for tracking the average of time-varying signals \cite{lns-v.85}. Indeed, these methods only require sparse, yet connected, networks. In Figure \ref{fig:overall_architecture} we show the flow of information for the proposed coordinated controller. 
Each controller in Figure \ref{fig:v_restoring} is a PI regulator with an anti-windup scheme accounting for saturations on $\Delta V_{PoL}$ and $\Delta\phi_{PoL}$. These saturations are needed to limit the amplitude and phase deviations, thus preventing the opening of breakers connecting inverters to the PoL. When saturations are not active, $\Delta V_{PoL}$ and $\Delta\phi_{PoL}$ are computed through the standard PI formulae
\begin{small}
	\begin{align}
	\label{eq:deltaVPCCctrl}\Delta V_{PoL}(t) &= K_{PV}(V_{PoL}^*-V_{PoL}(t))+K_{IV}\int_0^t(V_{PoL}^*-V_{PoL}(\tau))~d\tau\\
	\label{eq:deltaphictrl}			\Delta\phi_{PoL}(t) &= K_{P\phi}(0-\phi_{PoL}(t))+K_{I\phi}\int_0^t(0-\phi_{PoL}(\tau))~d\tau.
	\end{align}
\end{small}
Next, we discuss the tuning of PI parameters. The integral time constants $T_{IV}=K_{PV}/K_{IV}$ and $T_{I\phi}=K_{P\phi}/K_{I\phi}$ in Figure \ref{fig:v_restoring} are chosen to make the corresponding PI control loop sufficiently slower than the inner PnP control loop. To this aim, we use the following first-order approximation of each DGU equipped with the corresponding stabilizing PnP controller
\begin{equation*}
\begin{aligned}
V^d_{i} = \frac{1}{1+sT^d_{i}}V^{*d}_{i}\qquad	V^q_{i} = \frac{1}{1+sT^q_{i}}V^{*q}_{i}.
\end{aligned}
\end{equation*}	
Furthermore, since the overall PnP architecture is stable and due to the presence of output impedance for each DGU, we can also state that
\begin{equation*}
\begin{aligned}
V^d_{PoL,i} = \frac{\mu^d_{i}e^{-s\tau^d_{i}}}{1+sT^d_{i}}V^{*d}_{i}\qquad	V^q_{PoL,i} = \frac{\mu^q_{i}e^{-s\tau^q_{i}}}{1+sT^q_{i}}V^{*q}_{i}
\end{aligned}
\end{equation*}
where $\mu^d_{i}$, $\mu^q_{i}$, $\tau^d_{i}$ and $\tau^q_{i}$ depend on the output impedance. Using \eqref{eq:pcci}, assuming $\tau_i=\tau^d_{i}\approx\tau^q_{i}$, $T_i=T^d_{i}\approx T^q_{i}$ and setting $\mu_{i}=\sqrt{(\mu^d_{i})^2+(\mu^q_{i})^2}$, we obtain
\begin{equation}
\label{eq:PCCifun}
\begin{aligned}
V_{PoL,i} = \frac{\mu_{i}e^{-s\tau_i}}{1+sT_{i}}V_i^*\qquad\phi_{PoL,i} = \phi_i^*.
\end{aligned}
\end{equation}			
Therefore, using \eqref{eq:PCCifun} and \eqref{eq:averagesPCC}, we can provide a linear model of the effect of $V_i^*$ and $\phi_i^*$ on $V_{PoL}$ and $\phi_{PoL}$, respectively. Furthermore, since $V_i^* = V_{PoL}^*+\Delta V_{PoL}$ and $\phi_i^*=0+\Delta\phi_{PoL}$, we also obtain a small-signal model of the effect of $\Delta V_{PoL}$ and $\Delta\phi_{PoL}$ on $V_{PoL}$ and $\phi_{PoL}$, respectively.\\
As regards the phase deviation, since $\phi_{PoL,i} = \phi_i^* = \Delta\phi_{PoL}$, we can easily derive a simplified model as
$$
\phi_{PoL}=\frac{\Delta\phi_{PoL}}{N}
$$
which can be used for computing the PI control action in \eqref{eq:deltaphictrl}.\\
For deriving a simplified model of the amplitude deviation, the closed-loop DGU dynamics in the first control layer must be considered. The second control layer should act mostly when the first control layer is at steady-state. Hence, using \eqref{eq:PCCifun}, we can write a local approximate dynamics as 
$$
V_{PCC,i} = \frac{\mu_{i}e^{-s\tau_{PCC}}}{1+sT_{PCC}}V_i^*=\frac{\mu_{i}e^{-s\tau_{PCC}}}{1+sT_{PCC}}(V_{PCC}^*+\Delta V_{PCC})
$$
where $\tau_{PoL}=\max (\tau_1,\ldots,\tau_N)$ and $T_{PoL}=\max (T_1,\ldots,T_N)$. Next, using \eqref{eq:averagesPCC}, we can derive
$$
V_{PoL} = \frac{\mu e^{-s\tau_{PoL}}}{1+sT_{PoL}}V_{PoL}^*+\frac{\mu e^{-s\tau_{PoL}}}{1+sT_{PoL}}\Delta V_{PoL}
$$
where $\mu = \sum_{i=1}^N \frac{\mu_i}{N}$. In conclusion,  we tune the gains of the PI controller assuming the system under control has the transfer function
$$
\frac{V_{PoL}}{\Delta V_{PoL}}=\frac{\mu e^{-s\tau_{PoL}}}{1+sT_{PoL}}.
$$

\subsection{Sharing of reactive power}
\label{sec:sharingQ}		
\begin{figure}[!htb]
	\centering
	\includegraphics[scale=0.42]{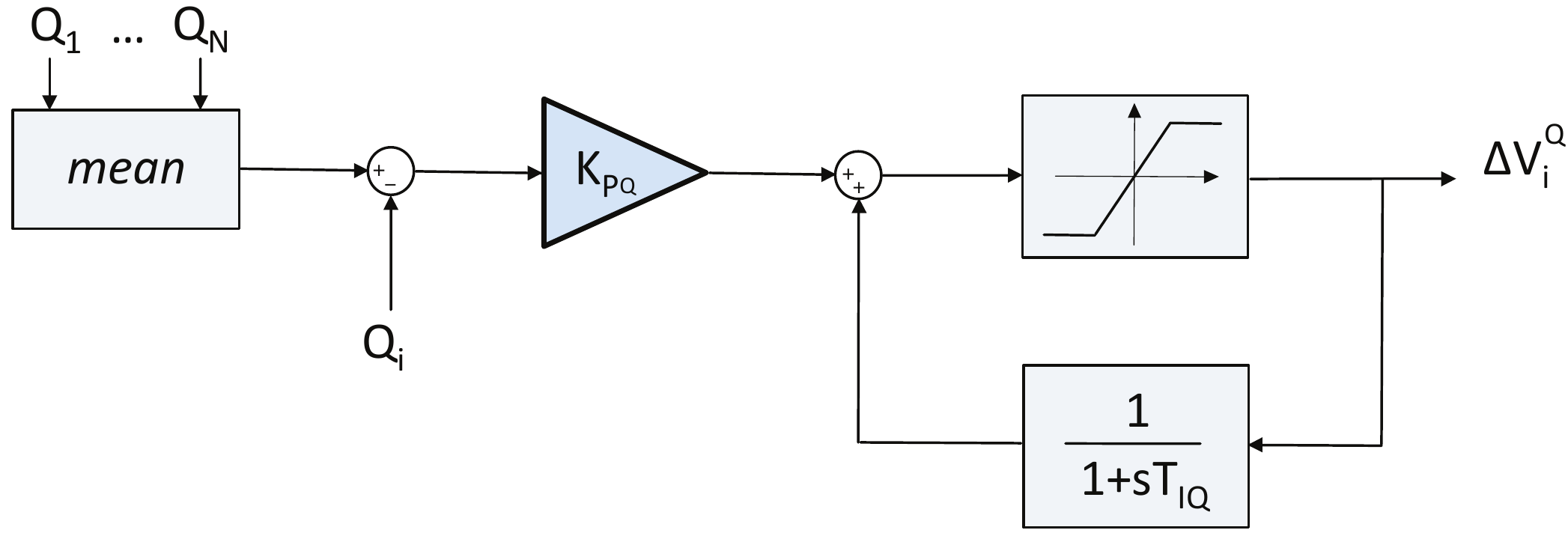}
	\caption{Control scheme for the computation of $\Delta V^Q_i$. Parameters $K_{PQ}$ and $T_{IQ}$ are the reactive power proportional term and the reactive power time constant, respectively.}
	\label{fig:q_restoring}
\end{figure}

Inner PnP regulators complemented with coordinated controllers for voltage tracking at the PoL cannot alone guarantee accurate reactive power sharing among DGUs. For this reason, we propose an additional coordinated controller dedicated to this aim. We assume that DGUs are connected to the PoL through mostly inductive lines\footnote{Results can be easily adapted to the case of lines that are mostly resistive.}. In this case,  the sharing of the reactive power is due to the amplitude of the voltages \cite{Yazdani2010,Teodorescu2011}. 
Therefore, we propose to change the amplitude of the set-point for each DGU as 
\begin{equation}
V_i^* = V_{PoL}^* + \Delta V_{PoL} + \Delta V^Q_{i},
\label{eq:set_point_Q}
\end{equation}
where $V_{PoL}^*$ is the reference for voltage at PoL, $\Delta V_{PoL}$ is computed as in \eqref{eq:deltaVPCCctrl} and $\Delta V^Q_{i}$ is a voltage correction to guarantee reactive power sharing. Voltage $\Delta V^Q_{i}$ is computed by the PI controller equipped with anti-windup shown in Figure \ref{fig:q_restoring}. In particular, when the saturation on $\Delta V^Q_{i}$ is not active, one has  
\begin{small}
	\begin{equation*}
	\Delta V^Q_{i}(t) = K_{PQ}(Q(t)-Q_i(t))+K_{IQ}\int_0^t(Q(\tau)-Q_i(\tau))~d\tau
	\end{equation*}
\end{small}
where $Q_i$ is the reactive power injected by the local DGU and $Q(t)$ is the average of the injected reactive powers. The PI regulator in Figure \ref{fig:q_restoring} is replicated in each DGU. Moreover, the whole control layer requires the communication network displayed in Figure \ref{fig:overall_architecture} since all units must exchange the values of reactive power $Q_i(t)$ for computing locally average $Q(t)$.\\
Similarly to what we have done for the PI regulators in Figure \ref{fig:v_restoring}, each integral time constant $T_{IQ}$ is designed to make the corresponding PI controller slower than the inner PnP ones.

\section{Experimental results}
\label{sec:exp_res}		

\subsection{Microgrid setup}

\begin{figure}[!htb]
	\centering
	\begin{subfigure}[!htb]{0.35\textwidth}
	\includegraphics[width=.9\textwidth]{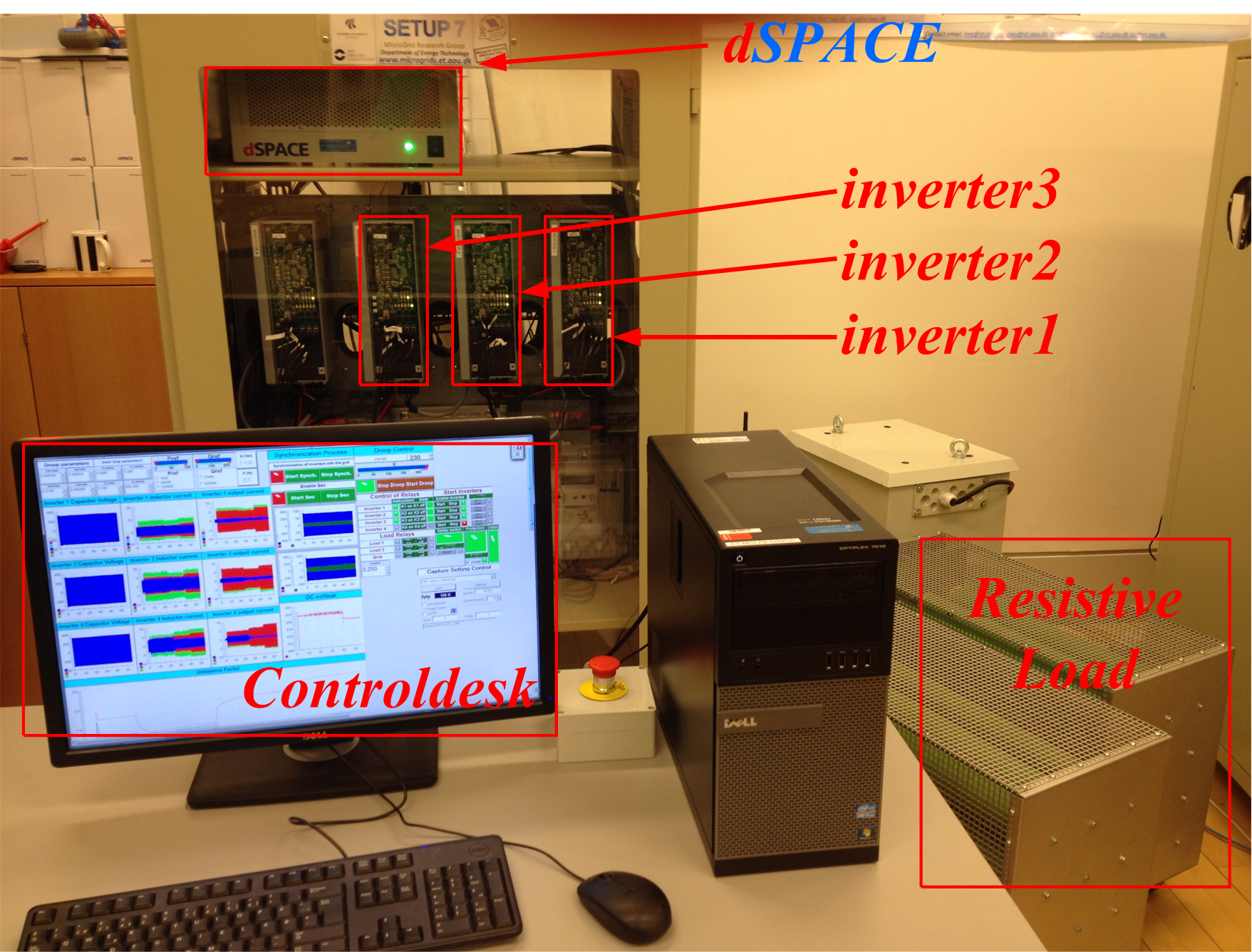}
	\caption{Experimental ImG setup.}
	\label{fig:exp_setup}
	\end{subfigure}
	\begin{subfigure}[!htb]{0.6\textwidth}
	\centering
	\includegraphics[width=1\textwidth]{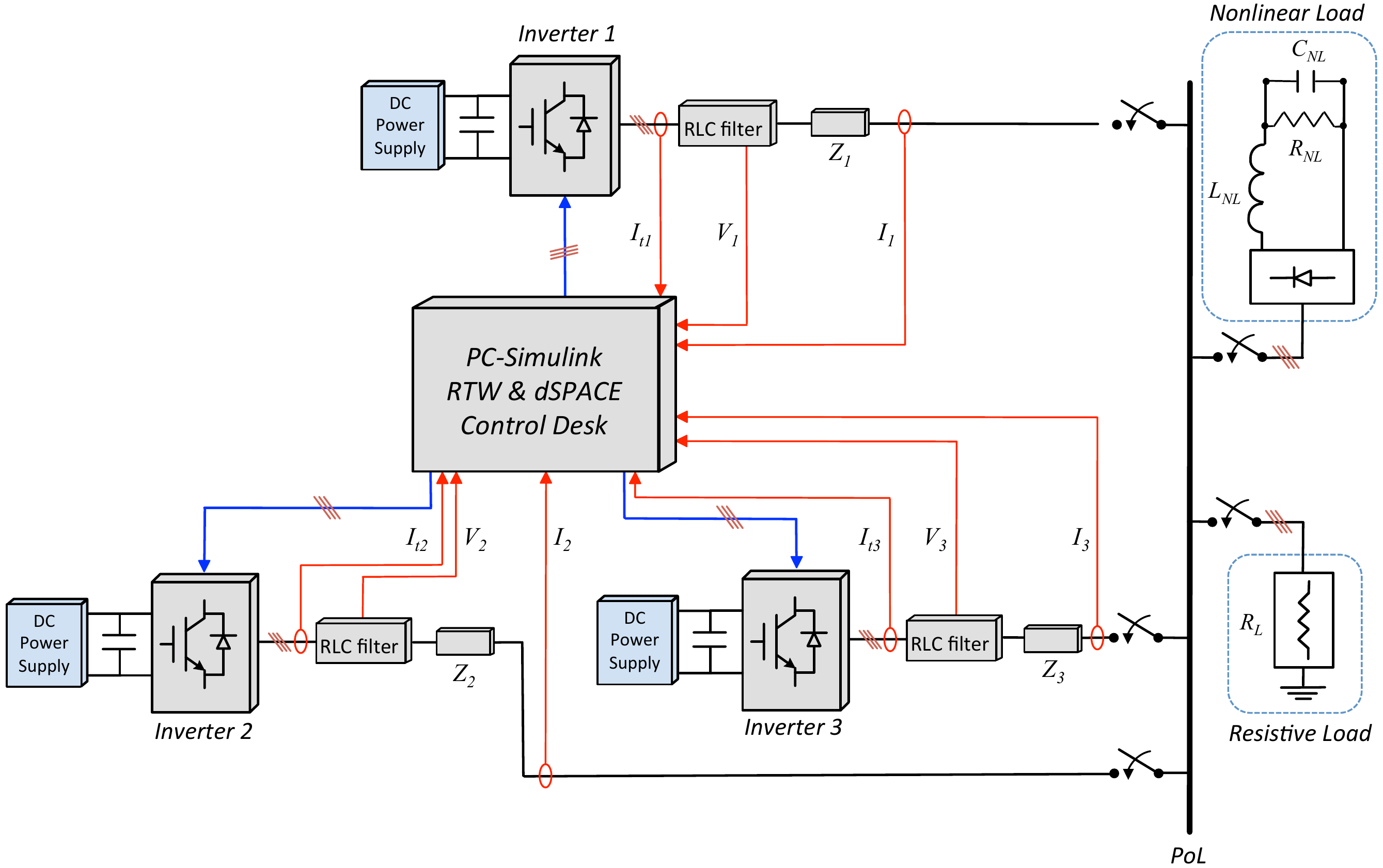}
	\caption{Scheme of the experimental setup. Red lines indicate the electrical variables measured at each DGU level. Blue lines represent the control commands to the inverters.}
	\label{fig:exp_setup_scheme}
\end{subfigure}
	\label{fig:setup}
	\caption{Experimental validation: ImG setup and implemented control scheme.}
\end{figure}

We tested the performance of the proposed approach using the ImG platform in Figure \ref{fig:exp_setup}; it consists of three \textit{Danfoss} inverters (2.2kVA with $RLC$ filters), a dSPACE1103 control board and LEM sensors. Inverters operate in parallel to emulate DGUs while different load conditions are obtained by connecting to the bus resistive loads and/or a diode rectifier. All the inverters are supplied by a DC source generator, therefore neither renewable sources nor energy storage devices are present in the experimental setup. Although this does not allow to study the effect of power fluctuations from renewable sources, the reliability of our experimental validation is guaranteed by the fact that, in general, changes in the power supplied by renewables take place at a timescale that is slower than the one we are interested in for stability analysis.

The controllers have been implemented in Simulink and compiled to the dSPACE system in order to command the inverter switches at a frequency of 10 kHz. Although the dSPACE platform is unique (see also Figure \ref{fig:exp_setup_scheme}), separate local controllers for each inverter were implemented so as to guarantee the control architecture can be implemented in a real distributed inverter system. The scheme of the experimental setup is depicted in Figure \ref{fig:exp_setup}.

In the performed experiments, we make sure that, at time $t = 0$ s, all the controllers are already activated so that all the voltages at the PCCs start from their reference value (230 V). 
 
The control and electrical parameters are reported in Table \ref{tbl:electrical_control_setup}.
Furthermore, we highlight that the clocks of local controllers have been synchronized using the procedure described in Section \ref{sec:synchronization_process}.

In the following sections, we validate primary PnP controllers under linear, unbalanced and nonlinear load conditions, as well as the combination of such primary layer with the proposed secondary coordinated controllers. 

\subsection{Experimental results: plug-and-play control}
\subsubsection{Voltage regulation at the PCCs with resistive load}
\label{sec:track_R_load}
In this first experiment, we test the capability of PnP primary controllers to handle connection and disconnection of inverters in a bus-connected ImG. At this stage, coordinated controllers described in Section \ref{sec:coordinatedControl} are not used. 

\begin{figure}[!htb]
	\centering
	\begin{subfigure}[!htb]{0.48\textwidth}
		\centering
		\includegraphics[width=1\textwidth]{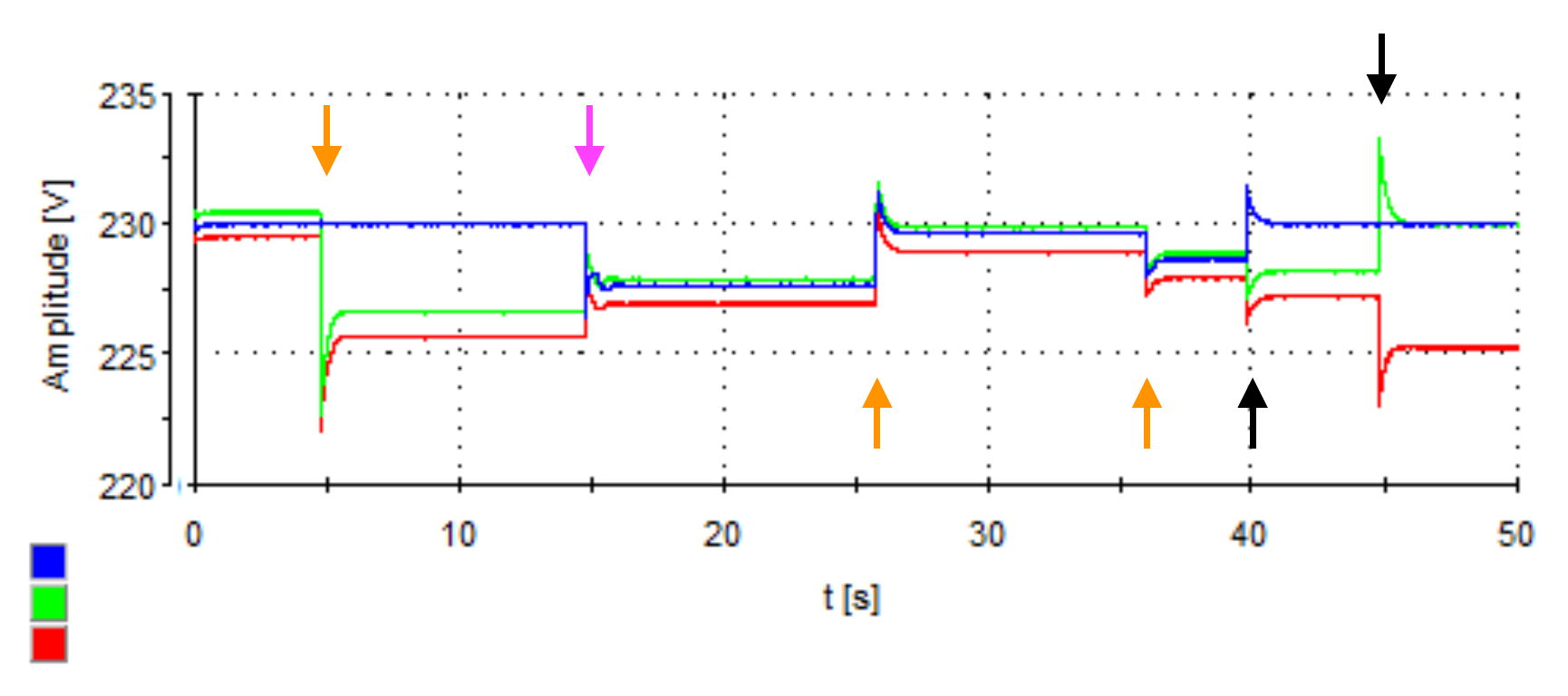}
		\caption{Voltages $V_{i}$ at the PCCs (phase $a$).}
		\label{fig:trackRload_voltage}
	\end{subfigure}
	\begin{subfigure}[!htb]{0.48\textwidth}
		\centering
		\includegraphics[width=1\textwidth]{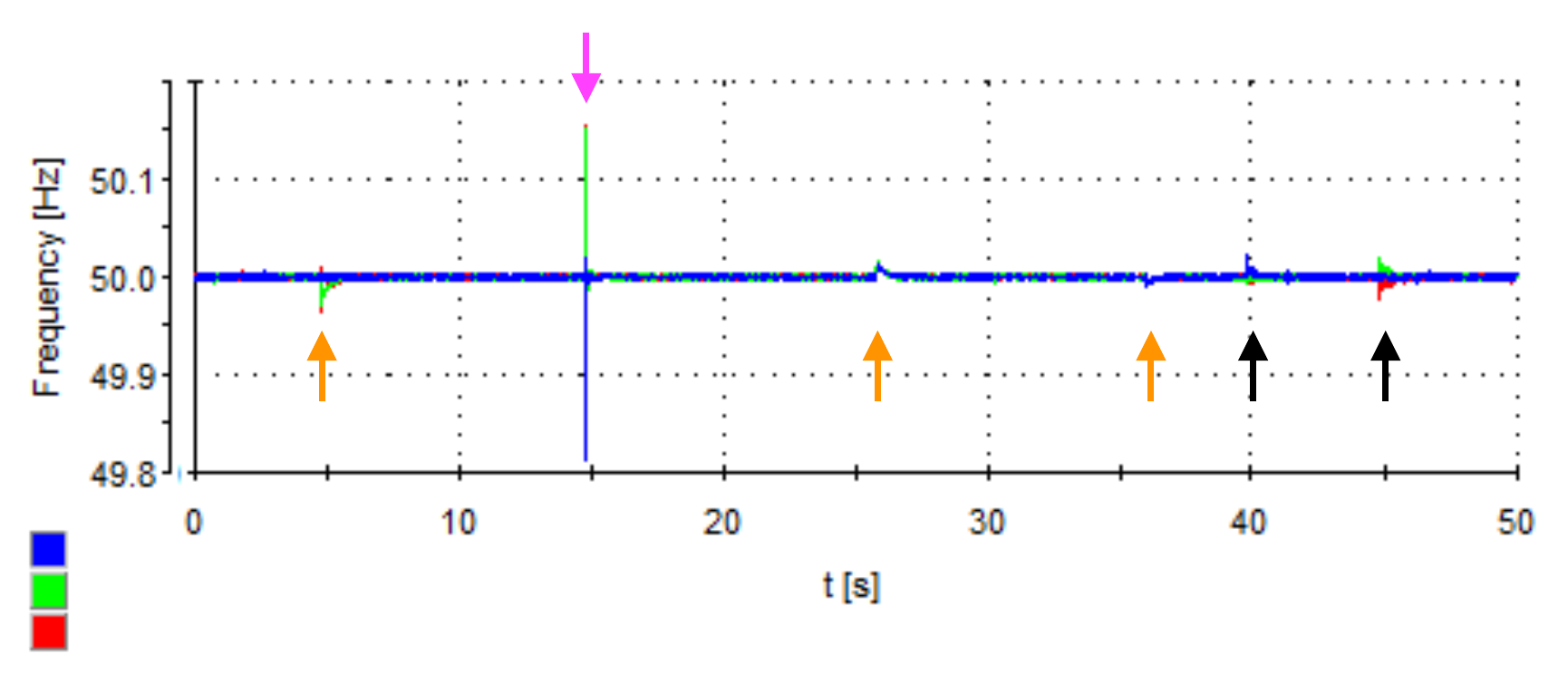}
		\caption{Frequencies (phase $a$).}
		\label{fig:trackRload_frequency}
	\end{subfigure}
	\begin{subfigure}[!htb]{0.48\textwidth}
		\centering
		\includegraphics[width=1\textwidth]{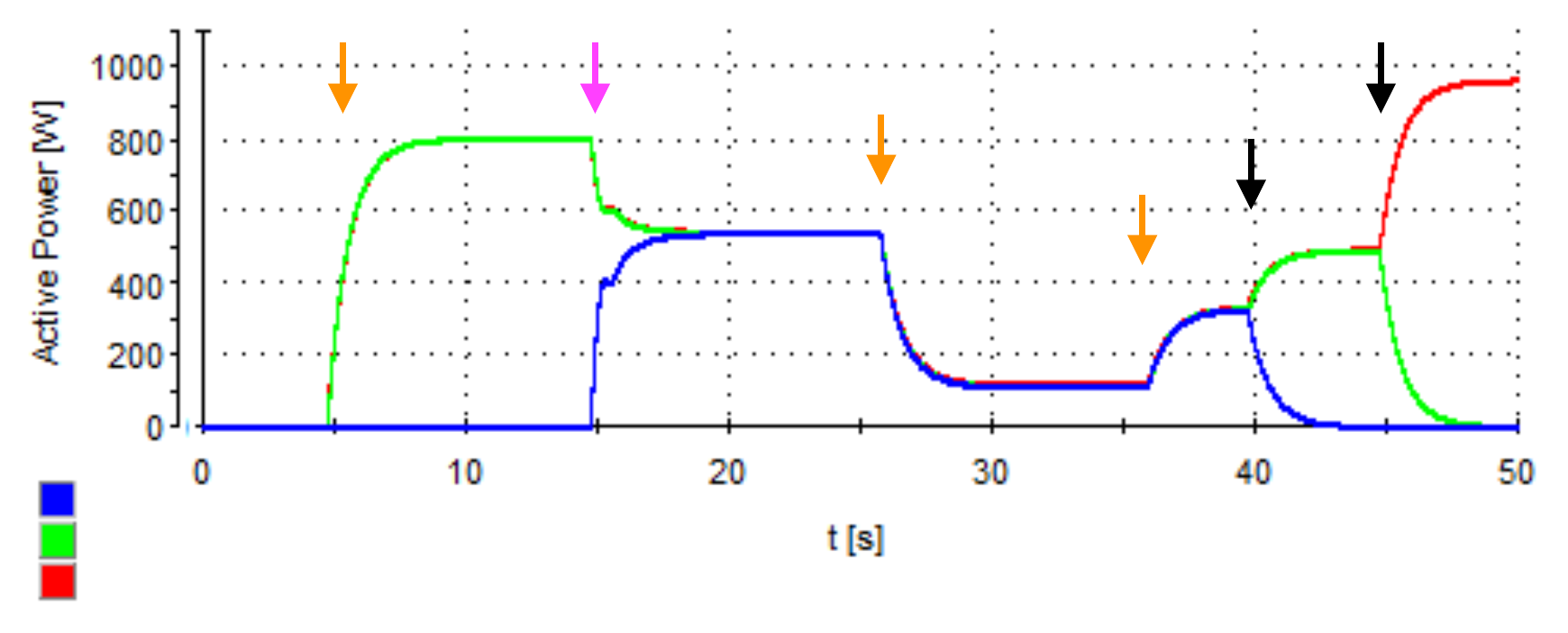}
		\caption{Active power provided by the inverters to the load.}
		\label{fig:trackRload_active}
	\end{subfigure}
	\begin{subfigure}[!htb]{0.48\textwidth}
		\includegraphics[width=1\textwidth]{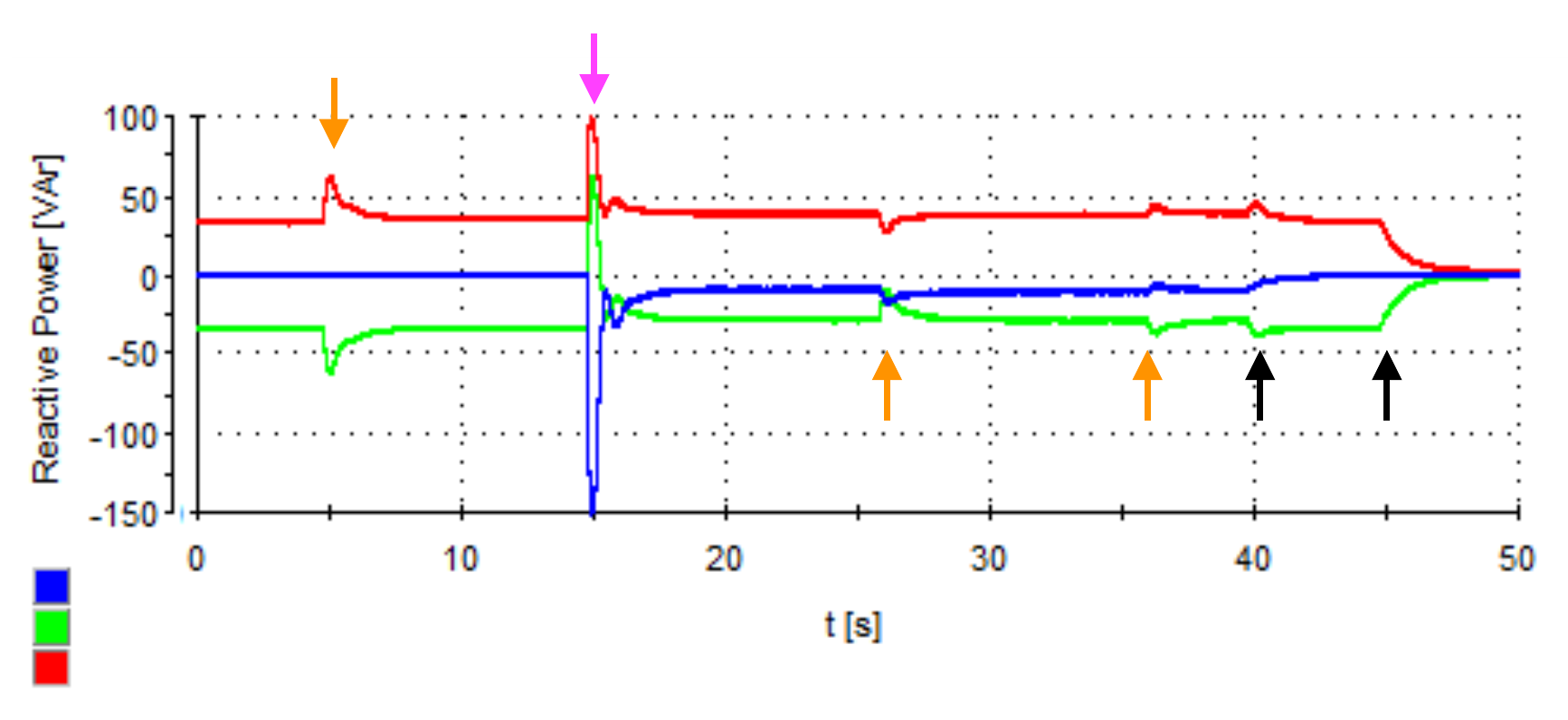}
		\caption{Reactive power provided by the inverters to the load.}
		\label{fig:trackRload_reactive}
	\end{subfigure}
	\caption{Voltage regulation at the PCCs with resistive load (Section \ref{sec:track_R_load}). Red, green and blue lines are, respectively, for VSC 1, 2 and 3. Load change, plug-in and unplugging events are indicated with orange, magenta and black arrows, respectively.}
	\label{fig:trackRload}
\end{figure}

At time $t=0$ s, the first and second inverter are connected to the bus; since there is no load at the PoL, the RMS voltages at PCCs 1 and 2 (red and green line, respectively, in Figure \ref{fig:trackRload_voltage}) coincide with the reference. 

At $t=5$ s, we connect a resistive load ($R=92$ $\Omega$) at the PoL. Consequently, the active powers provided by inverters 1 and 2 increase in order to compensate the load (see Figure \ref{fig:trackRload_active}). We also notice that the frequencies are promptly restored after the load connection (as shown in Figure \ref{fig:trackRload_frequency}). 

Since the voltage references at PCCs are fixed, PnP controllers alone cannot guarantee a good voltage regulation at the PoL. 
Moreover, as recalled in Section \ref{sec:ctrl_architecture}, the presence of a local virtual impedance induces a drop (proportional to the output current) in the corresponding reference voltage. This behavior is shown, for instance, in Figure \ref{fig:trackRload_voltage}, where we notice a decrement in the voltages at PCCs 1 and 2 when the load is connected to the PoL (at $t=5$ s).

At time $t=15$ s, we plug-in inverter 3. This event induces spikes in the frequencies (see Figure \ref{fig:trackRload_frequency}), whose maximal amplitude, however, is less than $0.2$ Hz. Moreover, Figure \ref{fig:trackRload_active} shows that all the inverters provide the same active power to compensate the load.

At times $t=25$ s and $t=35$ s, we change the load to $R=460$ $\Omega$ and $R=154$ $\Omega$, respectively. These events generate drops in the active power of more than $50\%$. Moreover, voltages and frequencies are instantaneously restored (see Figures \ref{fig:trackRload_voltage} and \ref{fig:trackRload_frequency}). Differently from droop-controllers, PnP controllers are not inertia-based and hence they are capable to provide faster transients. 
 
Finally, at $t=40$ s and $t=45$ s we plug-out inverters 3 and 2, respectively, thus eventually feeding the resistive load with inverter 1 only. Also in this case, the impact of the unplugging events on the frequency profile is minor.

\subsubsection{Voltage regulation at the PCCs with unbalanced load}
\label{sec:track_unbalanced}			
\begin{figure}[!htb]
	\centering
	\begin{subfigure}[!htb]{0.48\textwidth}
		\centering
		\includegraphics[width=1\textwidth]{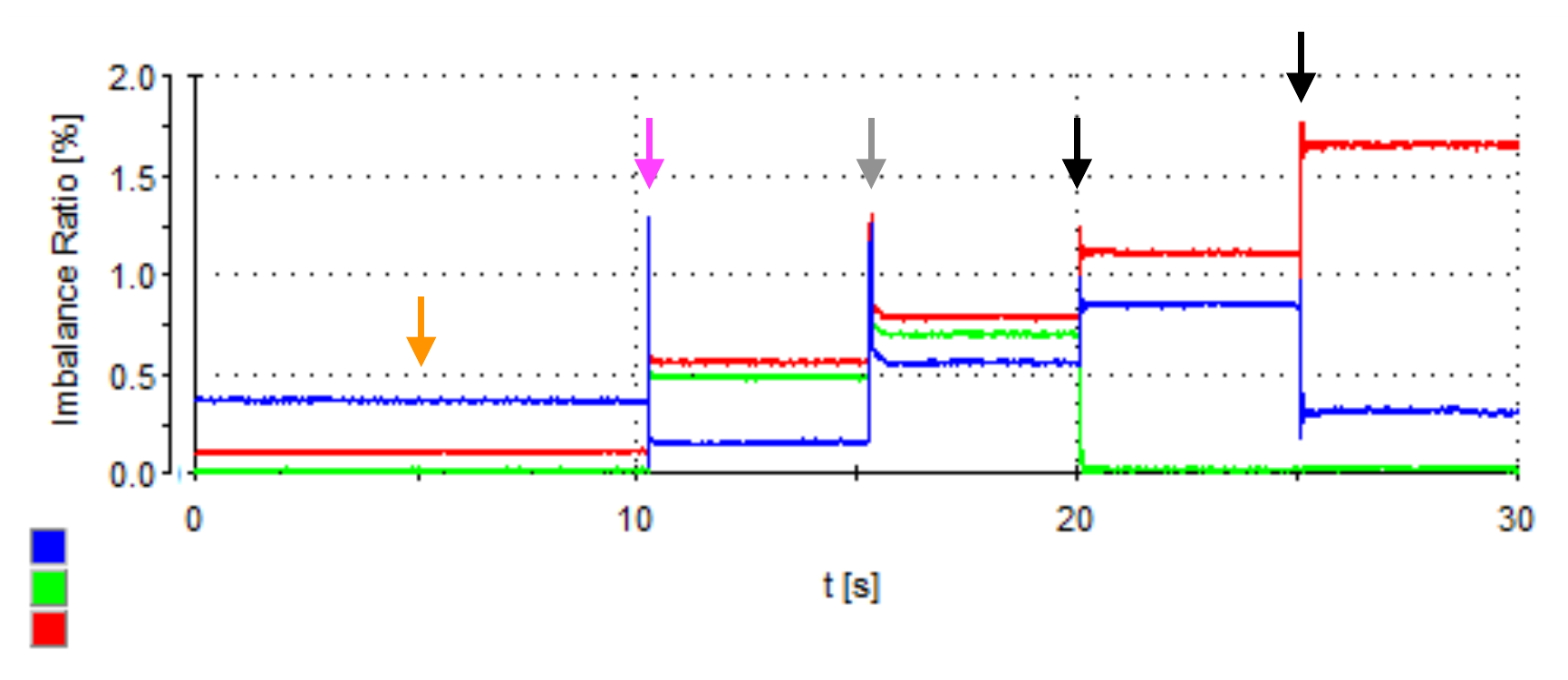}
		\caption{Imbalance ratios. Red, green and blue lines are, respectively, for VSC 1, 2 and 3. Load connection, phase $b$ load unbalancing, phase $c$ load unbalancing and unplugging events are indicated with orange, magenta, grey and black arrows, respectively.}
		\label{fig:trackUnbal_ratio}
	\end{subfigure}
	\begin{subfigure}[!htb]{0.48\textwidth}
		\centering
		\includegraphics[width=1\textwidth]{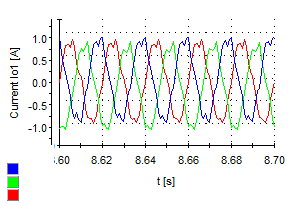}
		\caption{Output current for inverter 1 around time $t=8.65$ s (effect of balanced load connection at $t=5$ s).}
		\label{fig:trackUnbal_8s}
	\end{subfigure}
	\begin{subfigure}[!htb]{0.48\textwidth}
		\centering
		\includegraphics[width=1\textwidth]{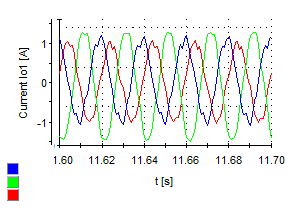}
		\caption{Output current for inverter 1 around time $t=11.65$ s (effect of load phase $b$ unbalancing at $t=10$ s).}
		\label{fig:trackUnbal_11s}
	\end{subfigure}
	\begin{subfigure}[!htb]{0.48\textwidth}
		\includegraphics[width=1\textwidth]{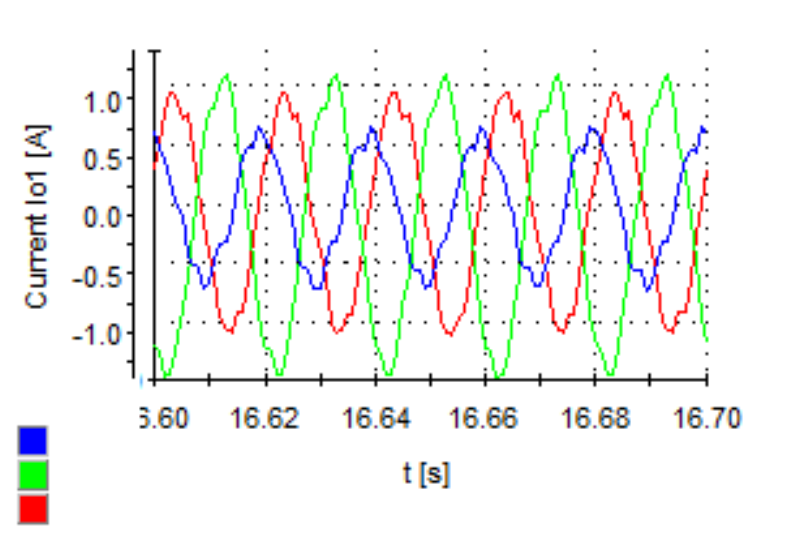}
		\caption{Output current for inverter 1 around time $t=16.65$ s (effect of load phase $c$ unbalancing at $t=15$ s).}
		\label{fig:trackUnbal_16s}
	\end{subfigure}
	\begin{subfigure}[!htb]{0.48\textwidth}
		\includegraphics[width=1.1\textwidth]{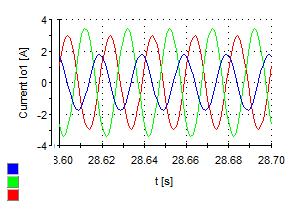}
		\caption{Output current for inverter 1 around time $t=28.65$ s (after the unplugging of VSCs 2 and 3).}
		\label{fig:trackUnbal_28s}
	\end{subfigure}
	\caption{Voltage regulation at the PCCs with unbalanced load (Section \ref{sec:track_unbalanced}).}
	\label{fig:trackUnbal}
\end{figure}

In this experiment, we show performance of PnP controllers under unbalanced load conditions. For the sake of simplicity, in Figures \ref{fig:trackUnbal_8s}-\ref{fig:trackUnbal_28s} we show the evolution of the main electrical quantities of inverter 1 only. 

At $t=0$ s, all the inverters are connected to the PoL and no load is present. Then, at $t=5$ s we connect a balanced resistive load ($R=115$ $\Omega$) to the common bus. Consequently, inverter 1 provides the output current shown in Figure \ref{fig:trackUnbal_8s}. At $t=10$ s, we change phase $b$ of the load to $R=57$ $\Omega$, thus causing the unbalance in the output current 1 shown in Figure \ref{fig:trackUnbal_11s}. Moreover, the average of the imbalanced ratios \cite{IEEE2009} for all the inverters is $0.5$ $\%$ (see Figure \ref{fig:trackUnbal_ratio}). At $t=15$ s, we change phase $c$ of the load to $R=230$ $\Omega$. As a consequence, in Figure \ref{fig:trackUnbal_16s} we note an additional unbalance in the output current 1. Moreover, from Figure \ref{fig:trackUnbal_ratio}, we see that the average of the imbalanced ratios increases to $0.75$ $\%$. Finally, at $t=20$ s and $t=25$ s, we unplug inverters 2 and 3, respectively: since inverter 1 must provide all the power required by the load, the amplitude of its output current increases (Figure \ref{fig:trackUnbal_28s}). Figure \ref{fig:trackUnbal_ratio} shows that also its imbalance ratio increases to $1$ $\%$ and then to $1.65$ $\%$. However, we notice that, during the whole experiment, the imbalance ratio is quite small and always lower than the maximum value (3 $\%$) recommended by IEEE in \cite{IEEE2009}.

\subsubsection{Voltage regulation at the PCCs with nonlinear load}	
\label{sec:pnpNL}
In this scenario, we show features of PnP controllers in presence of nonlinear loads. At time $t=0$ s, inverters 1 and 2 are connected to the diode rectifier shown in Figure \ref{fig:exp_setup_scheme}. Hence, the active power provided by inverter 3 is zero (see Figure \ref{fig:trackNLload_active_power}) and the THD{\footnote{For the sake of simplicity, in Figure \ref{fig:trackNLload_THD} we show only the THD indices of phase $a$ of the corresponding PCC voltages.} (Total Harmonic Distortion index in \cite{IEEE2009}) is higher for the voltages at PCCs 1 and 2 (as shown in Figure \ref{fig:trackNLload_THD}). At $t=5$ s, we increase the power required at the PoL by connecting a resistive load ($R=154$ $\Omega$) in parallel with the nonlinear one. 

The plugging-in operation of inverter 3 is performed at $t=15$ s. Notice that the frequencies are promptly restored to the nominal value (variations less than $0.2$Hz), total active power is equally shared between all inverters and THDs are reduced for all inverters. 

In order to assess the robustness of local PnP regulators to unknown load dynamics, at times $t = 25$ s and $t=35$ s, we switch the resistive load to $R = 460$ $\Omega$ and $R=154$ $\Omega$, respectively. Figures \ref{fig:trackNLload_voltage} and \ref{fig:trackNLload_frequency} show fast transients of voltages and frequencies at the PCCs.

Finally, at $t=40$ s and $t=45$ s, we unplug inverter 3 and 2, respectively. Consequently, the THD of the voltage at PCC 1 increases. However, as shown in Figure \ref{fig:trackNLload_THD}, the  THD values are always below the maximum limit ($5\%$) recommended in \cite{IEEE2009}. Concluding, this experiment reveals that, even in absence of inner resonant controllers, PnP regulators are capable to guarantee high levels of robustness to load variations and harmonic attenuation.	

\begin{figure}[!htb]
	\centering
	\begin{subfigure}[!htb]{0.48\textwidth}
		\centering
		\includegraphics[width=1\textwidth]{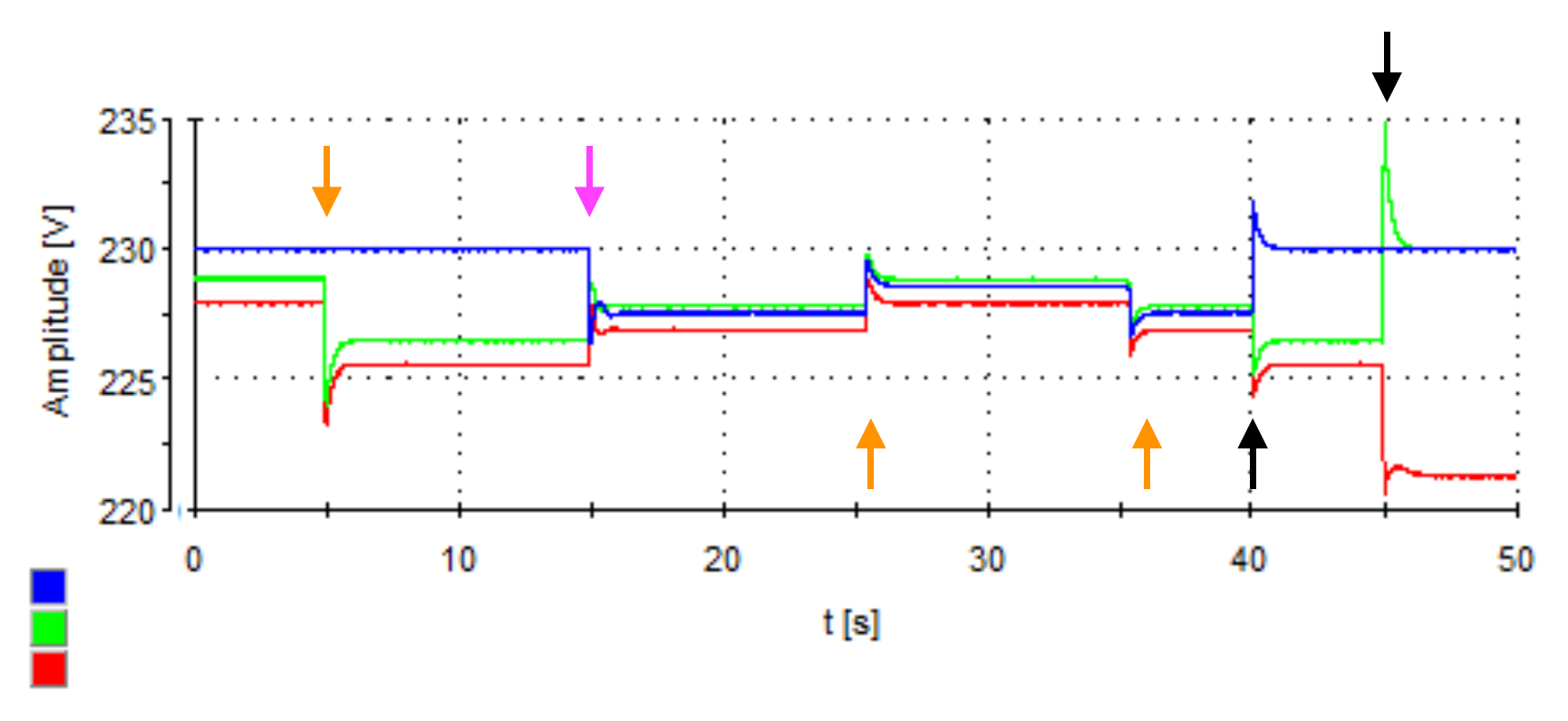}
		\caption{Voltages $V_{i}$ at the PCCs (phase $a$).}
		\label{fig:trackNLload_voltage}
	\end{subfigure}
	\begin{subfigure}[!htb]{0.48\textwidth}
		\centering
		\includegraphics[width=1\textwidth]{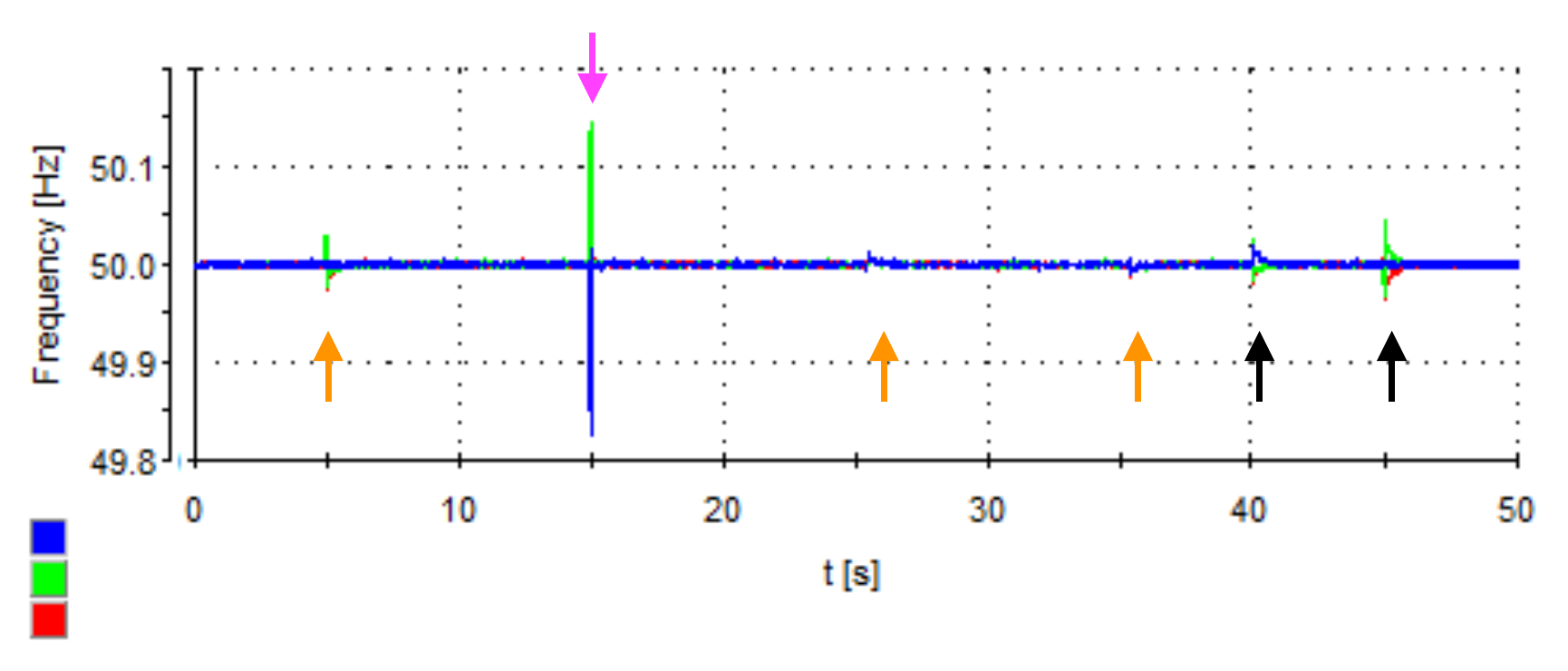}
		\caption{Frequencies (phase $a$).}
		\label{fig:trackNLload_frequency}
	\end{subfigure}
	\begin{subfigure}[!htb]{0.48\textwidth}
		\centering
		\includegraphics[width=1\textwidth]{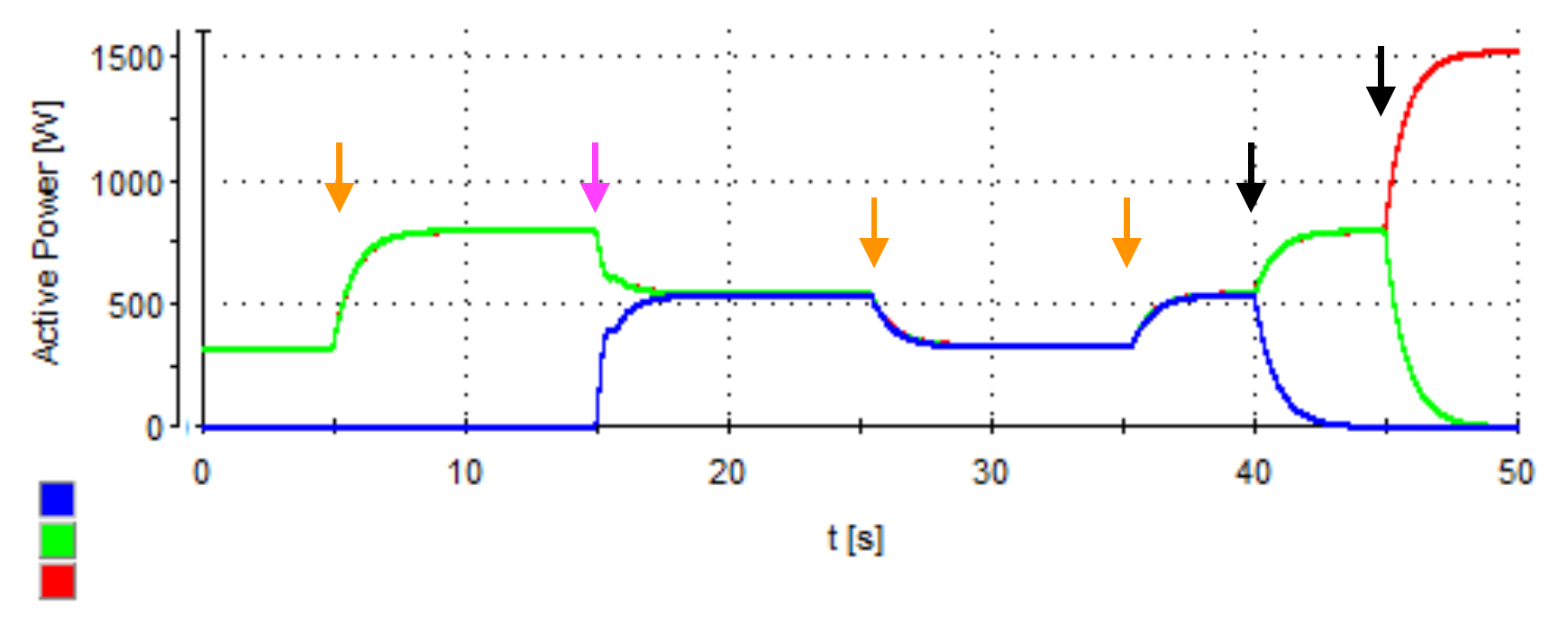}
		\caption{Active power provided by the inverters to the load.}
		\label{fig:trackNLload_active_power}
	\end{subfigure}
	\begin{subfigure}[!htb]{0.48\textwidth}
		\includegraphics[width=1\textwidth]{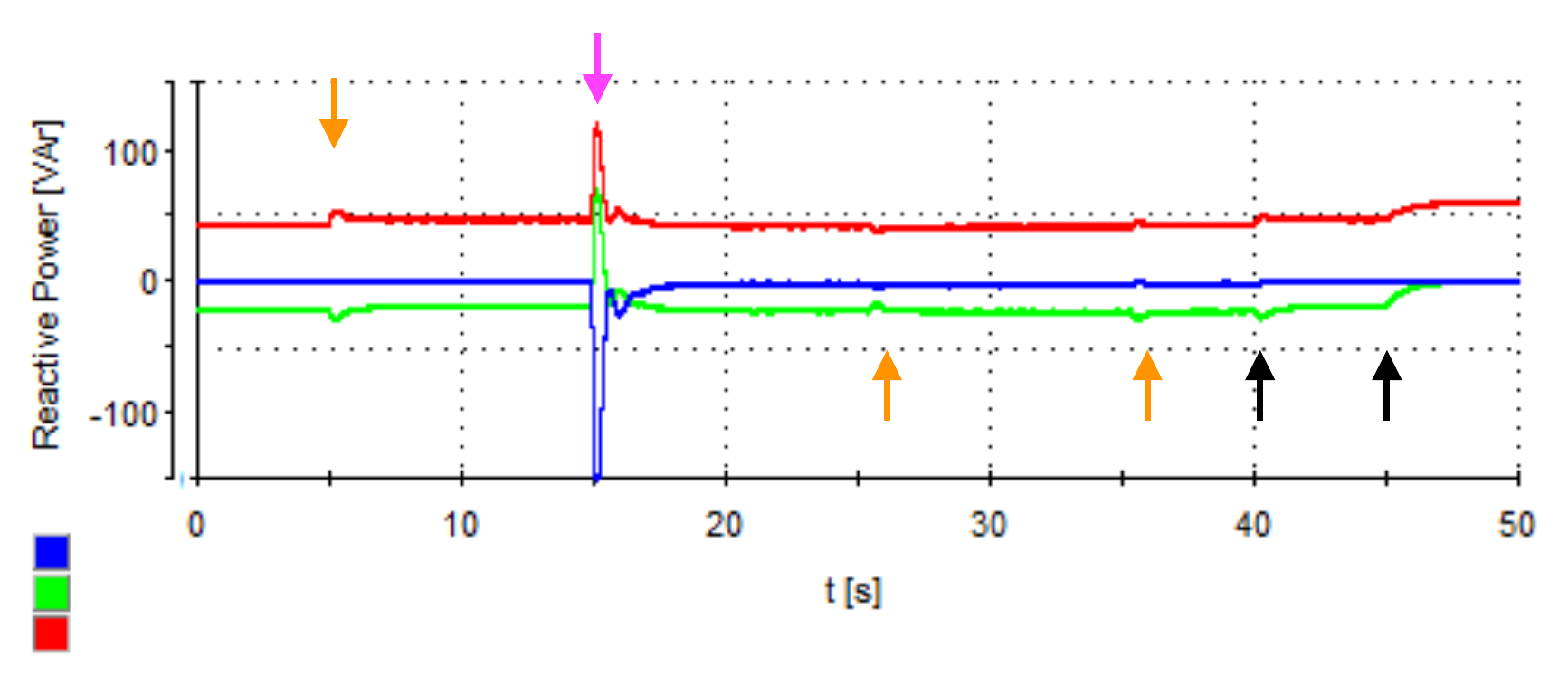}
		\caption{Reactive power provided by the inverters to the load.}
		\label{fig:trackNLload_reactive_power}
	\end{subfigure}
	\begin{subfigure}[!htb]{0.48\textwidth}
		\includegraphics[width=1\textwidth]{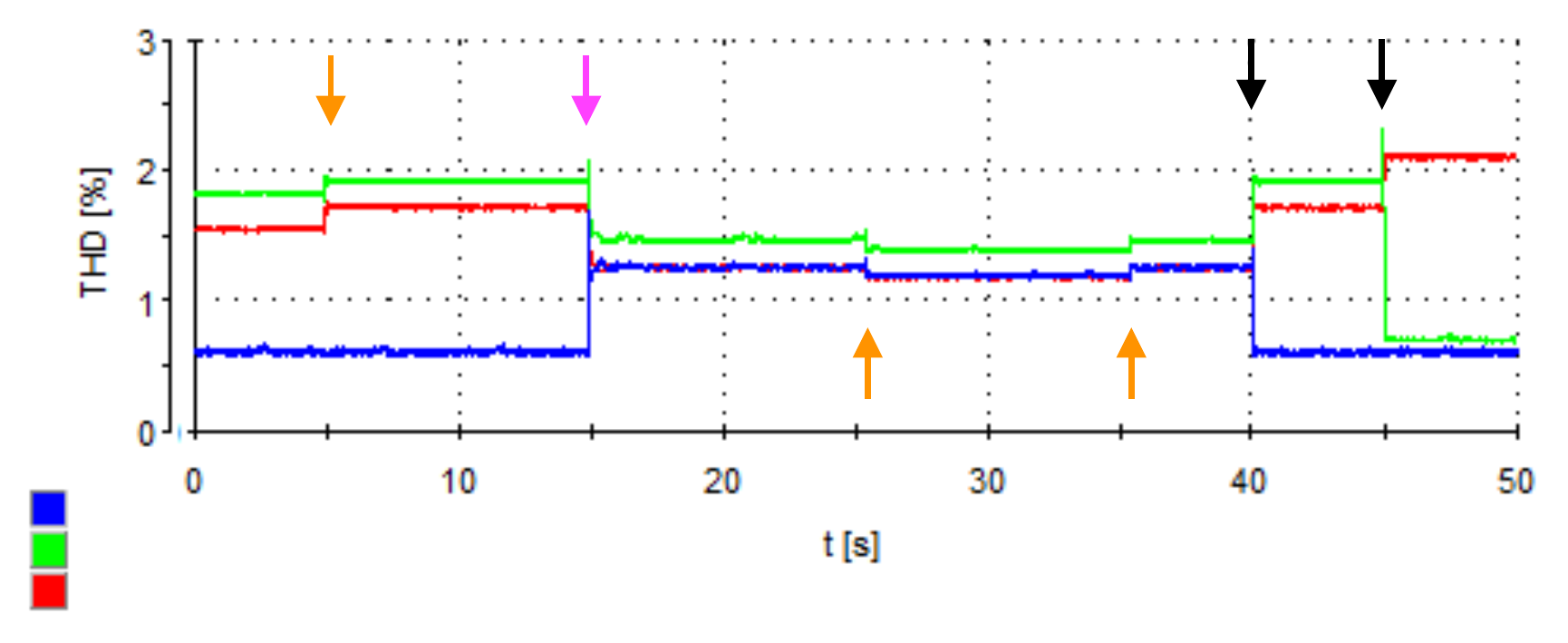}
		\caption{Total harmonic distortion.}
		\label{fig:trackNLload_THD}
	\end{subfigure}
	\caption{Voltage regulation at the PCCs with nonlinear load (Section \ref{sec:pnpNL}). Red, green and blue lines are, respectively, for VSC 1, 2 and 3. Load change, plug-in and unplugging events are indicated with orange, magenta and black arrows, respectively.}
	\label{fig:trackNLload}
\end{figure}

 \begin{table}[!htb]                 
	\centering
	\begin{tabular}{|l|c|c|c|}
		\hline
		Parameter & Symbol & Value & Units \\
		\hline
		\multicolumn{4}{|c|}{\textit{\textbf{Virtual impedance}}} \\  
		\hline
		
		\hline
		Virtual resistance& $R_{i,v}$ & 3 & $\Omega$\\
		\hline
		Virtual inductance& $L_{i,v}$ & 0.03 & H\\
		\hline
		\multicolumn{4}{|c|}{\textit{\textbf{Voltage tracking at the PoL}}} \\  
		\hline
		Module proportional term & $K_{PV} $ & $10^{-3}$ & -\\
		\hline
		Module integral term & $K_{IV} $ &  0.6 & -\\
		\hline
		Phase proportional term & $K_{P\phi} $ & $10^{-3}$ & -\\
		\hline
		Phase integral term & $K_{I\phi} $ & 4 & -\\
		\hline
		\multicolumn{4}{|c|}{\textit{\textbf{Reactive power sharing}}} \\  
		\hline
		Reactive power proportional term & $K_{PQ} $ &  $10^{-4}$ & - \\
		\hline
		Reactive power integral term & $K_{IQ} $ & $10^{-2}$ & -\\
		\hline
		\multicolumn{4}{|c|}{\textit{\textbf{Electrical Setup}}} \\  
		\hline
		PCC reference voltage & $V_{ref}$ & 230 & V \\
		\hline
		ImG frequency & $f_0$& 50 & Hz\\
		\hline
		Switching frequency & $f_{sw}$ & 10 & kHz \\
		\hline
		Filter resistance & $R_{ti}$ & 0.1 & $\Omega$\\
		\hline
		Filter inductance & $L_{ti}$ & 1.8 & mH\\
		\hline
		Filter capacitance & $C_{ti}$ & 25 & $\mu$F\\
		\hline
		Line resistance & $R_{i}$ & 0.1 & $\Omega$\\
		\hline
		Line inductance & $L_{i}$ & 1.8 & mH\\
		\hline
		Phase resistive load & $R_{a,b,c}$& 57-115-230-460& $\Omega$ \\
		\hline
		Nonlinear load &$R_{NL}$ & 460 & $\Omega$\\
		\hline
		
	\end{tabular}
	
	\caption{Electrical setup and control parameters.}	
	\label{tbl:electrical_control_setup}
\end{table}

\subsection{Experimental results: plug-and-play and coordinated control}
\label{sec:pnp_and_coordinated}
In this Section, we validate the combination of the coordinated control layer presented in Section \ref{sec:coordinatedControl} and PnP primary regulators.

As a first test, we consider the same experimental scenario as in Section \ref{sec:pnpNL}, complemented with the coordinated controllers presented in Section \ref{sec:restoringPCC} for voltage tracking at the PoL. This operation is desirable because, as shown in Figure \ref{fig:trackNLload_voltage}, PnP regulators alone fail to keep the PoL voltage at the nominal value, even though the voltages at the PCCs are stabilized. As highlighted in Section \ref{sec:coordinatedControl}, this issue is due to the fixed voltage set-points for the PnP controllers. In this experiment, we activate coordinated controllers at time $t=20$ s. Consequently, as shown in Figure \ref{fig:trackNLload_coord_voltage}, the voltages at the PCCs increase in order to track the nominal PoL voltage. Moreover, the proposed coordinated controllers are capable to keep the PoL voltage at the desired level (by  regulating the voltages at the PCCs) even when load changes (at times $t=25$ s and $t=35$ s) and disconnection of inverters (at $t=40$ s and $t=45$ s) are performed (see Figure \ref{fig:trackNLload_coord}). We also highlight that the presence of the coordinated controllers does not affect the frequency profiles (see Figure \ref{fig:trackNLload_coord_frequency}). On the other hand, Figure \ref{fig:trackNLload_coord_reactive_power} reveals that the total reactive power is still not shared equally between the inverters.	
\begin{figure}[!htb]
	\centering
	\begin{subfigure}[!htb]{0.48\textwidth}
		\centering
		\includegraphics[width=1\textwidth]{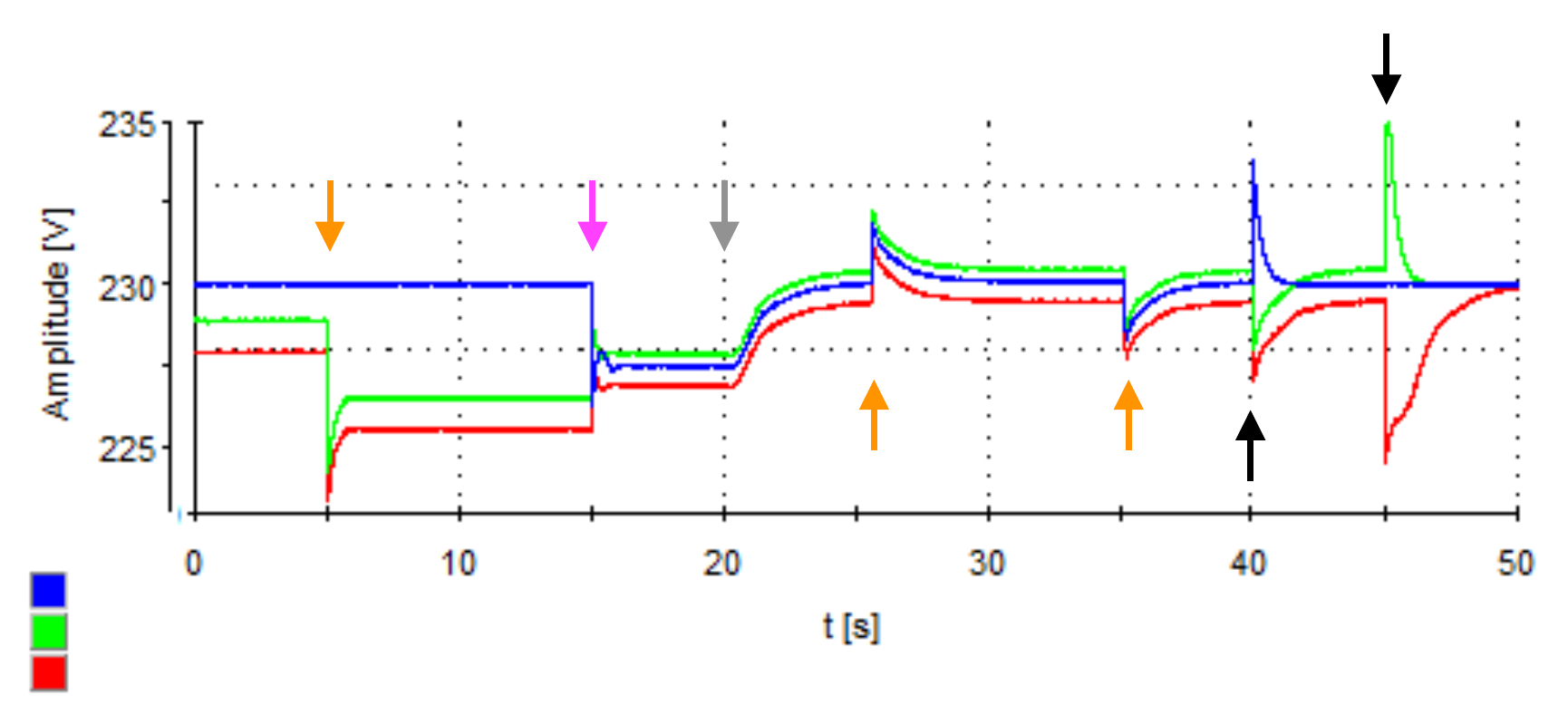}
		\caption{Voltages $V_{i}$ at the PCCs (phase $a$).}
		\label{fig:trackNLload_coord_voltage}
	\end{subfigure}
	\begin{subfigure}[!htb]{0.48\textwidth}
		\centering
		\includegraphics[width=1\textwidth]{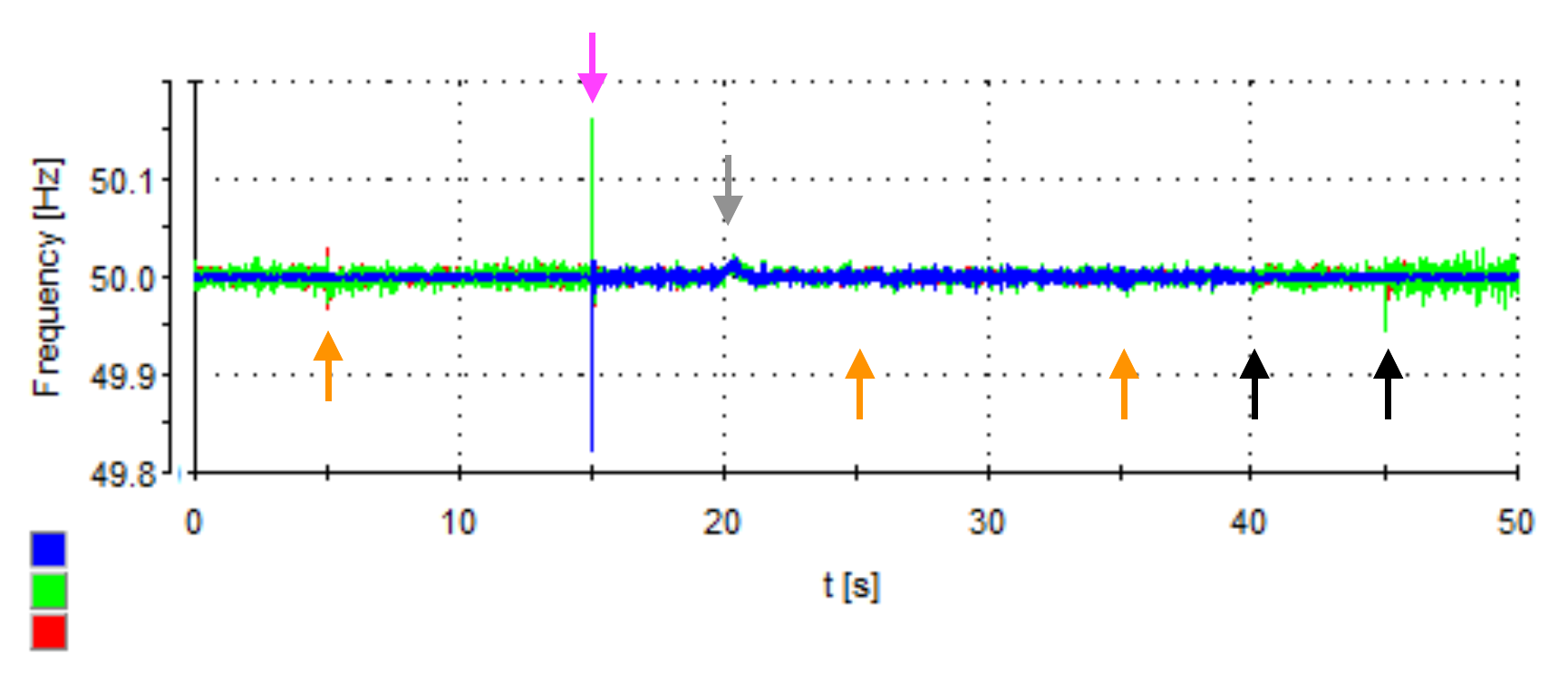}
		\caption{Frequencies (phase $a$).}
		\label{fig:trackNLload_coord_frequency}
	\end{subfigure}
	\begin{subfigure}[!htb]{0.48\textwidth}
		\centering
		\includegraphics[width=1\textwidth]{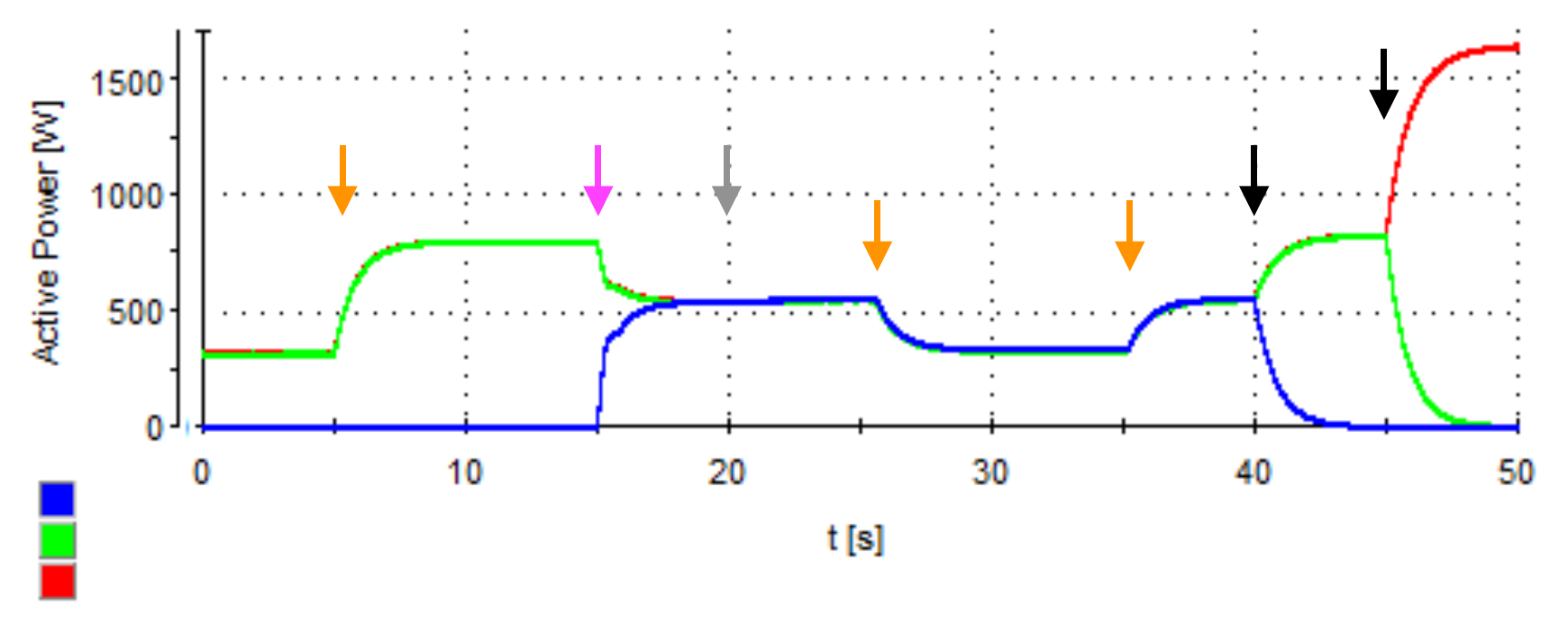}
		\caption{Active power provided by the inverters to the load.}
		\label{fig:trackNLload_coord_active_power}
	\end{subfigure}
	\begin{subfigure}[!htb]{0.48\textwidth}
		\includegraphics[width=1\textwidth]{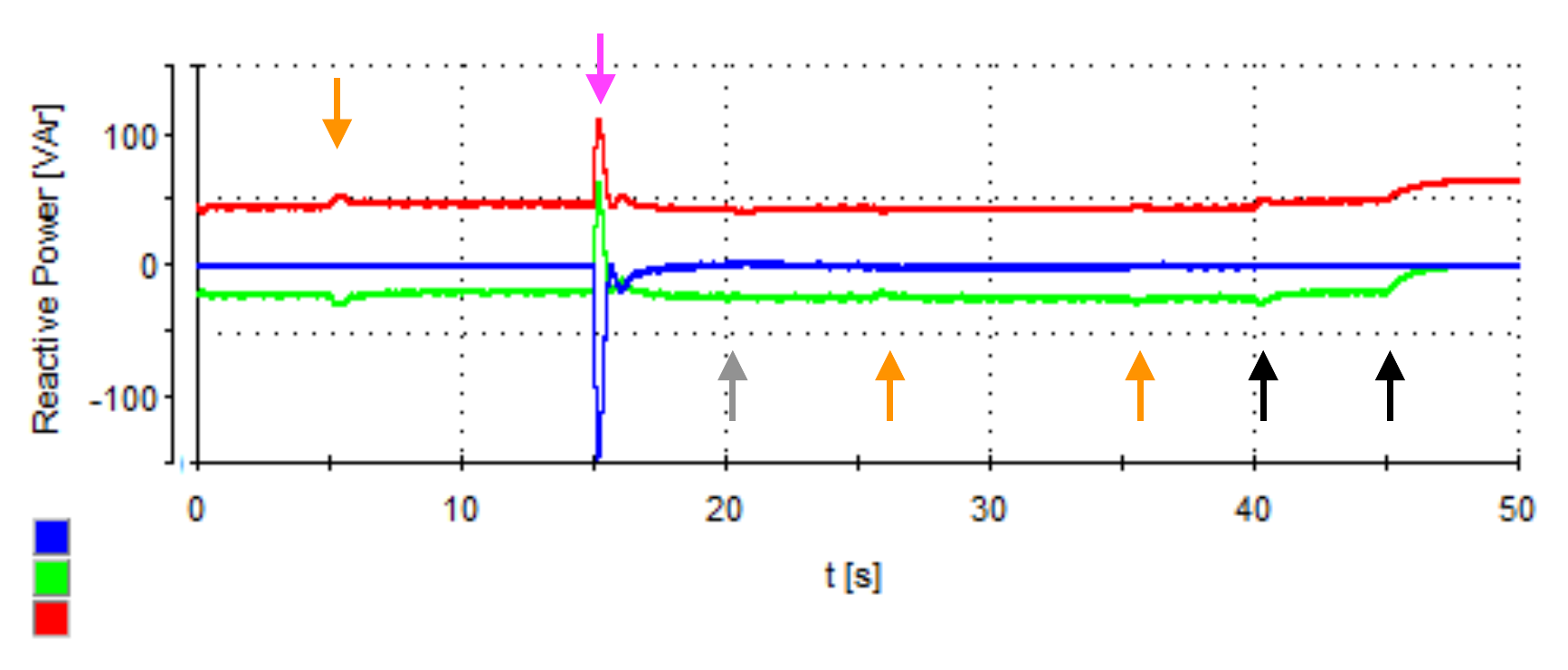}
		\caption{Reactive power provided by the inverters to the load.}
		\label{fig:trackNLload_coord_reactive_power}
	\end{subfigure}
	\caption{PnP regulators and coordinated controllers for voltage tracking at the PoL with nonlinear load. Red, green and blue lines are, respectively, for VSC 1, 2 and 3. Load change, plug-in and unplugging events are indicated with orange, magenta and black arrows, respectively. Moreover, the grey arrow denotes the activation of secondary controllers described in Section \ref{sec:restoringPCC}.}
	\label{fig:trackNLload_coord}
\end{figure}

In order to ameliorate also aspect, we run a second experiment in which the previous control scheme is complemented with coordinated controllers for reactive power sharing (described in Section \ref{sec:sharingQ}). In particular, we notice that, right after their activation (at time $t=20$ s), the total reactive power is equally shared (see Figure \ref{fig:trackNLload_coord_reactive_power_Q}, to be compared with Figure \ref{fig:trackNLload_coord_reactive_power}). This goal is achieved through the computation of terms $\Delta V_i^Q$ in \eqref{eq:set_point_Q}, which are shown in Figure \ref{fig:trackNLload_coord_deltaVrefQ}.

\begin{figure}[htb]
	\centering
	\begin{subfigure}[!htb]{0.48\textwidth}
		\centering
		\includegraphics[width=1\textwidth]{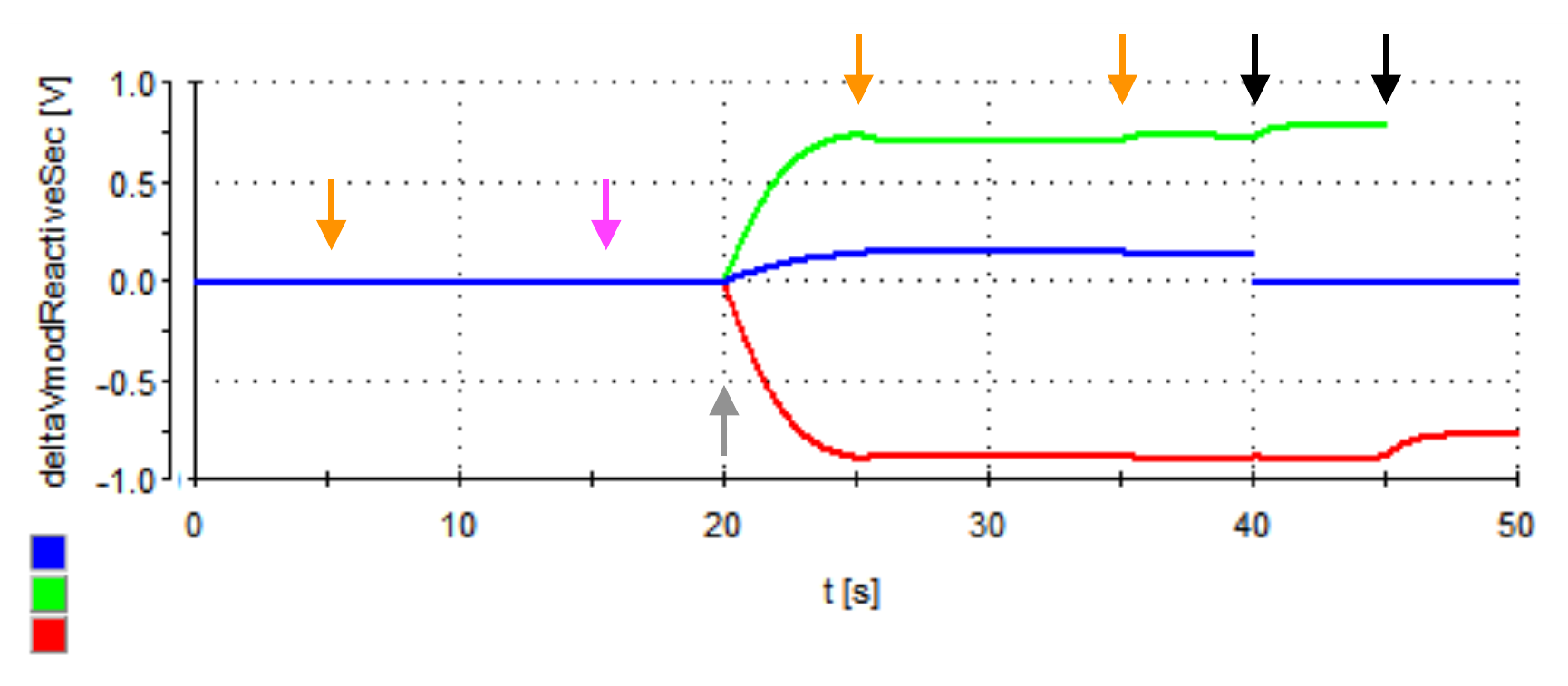}
		\caption{Amplitude deviations $\Delta V_i^Q$ in \eqref{eq:set_point_Q} leading to the sharing of total reactive power.}
		\label{fig:trackNLload_coord_deltaVrefQ}
	\end{subfigure}
	\begin{subfigure}[!htb]{0.48\textwidth}
		\includegraphics[width=1\textwidth]{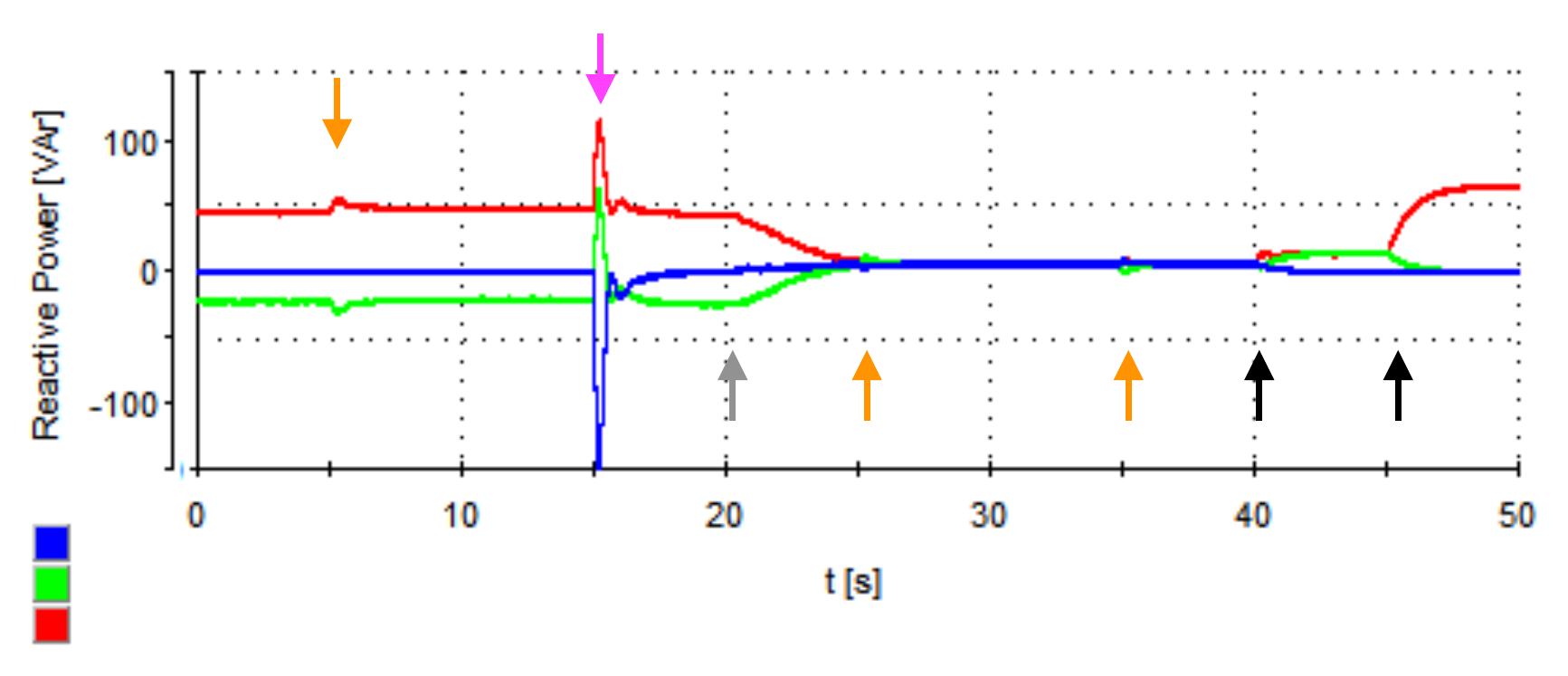}
		\caption{Reactive power produced by each VSC.}
		\label{fig:trackNLload_coord_reactive_power_Q}
	\end{subfigure}
	\caption{PnP regulators and coordinated controllers for voltage tracking at the PoL and reactive power sharing with nonlinear load. Red, green and blue lines are respectively for VSC 1, 2 and 3. Load change, plug-in and unplugging events are indicated with orange, magenta and black arrows, respectively. Moreover, the grey arrow denotes the simultaneous activation of secondary controllers described in Sections \ref{sec:restoringPCC} and \ref{sec:sharingQ}.}
	\label{fig:trackNLload_coord_Q}
\end{figure}

\section{Conclusions}
\label{sec:conclusions}
In this paper, we showed how to adapt the PnP control scheme \cite{riverso2015plug, Riverso2014c} for voltage and frequency regulation in AC ImGs to bus-connected topologies. Moreover, we introduced a secondary control layer and performed an experimental validation of the overall control hierarchy. Using the parallel connection of three VSCs, we showed that stability, efficient tracking of the voltage bus and sharing of reactive power can be obtained. Moreover, experiments with linear and nonlinear loads show that the harmonic distortion and imbalance ratio are kept within acceptable bounds. In our implementation, time-synchronization of DGU clocks and the secondary control layer assume all-to-all communication among DGUs. Future research will focus on distributing these computations as well, by exploiting only partial communication among DGUs and, as in \cite{ShafieeDistributed2014}, consensus-like protocols for estimating averages of global variables

\section{Acknowledgment}

The authors wish to thank Alessandro Floriduz for insightful discussions.

       \clearpage

\bibliographystyle{IEEEtran}
\bibliography{PnP_microgrids-AE}


%
%
%
\end{document}